\documentclass{ifacconf}

\usepackage{graphicx}      
\usepackage{natbib}        
\usepackage{subcaption}
\usepackage{amsmath,amssymb,amsfonts}
\allowdisplaybreaks

\begin{document}
\begin{frontmatter}

\title{Sparse Resource Allocation for Spreading Processes on Temporal-Switching Networks} 


\author[First]{Vera L. J. Somers} 
\author[Second]{Ian R. Manchester} 

\address[First]{ARC Training Centre in Optimisation Technologies, Integrated Methodologies, and Applications (OPTIMA), University of Melbourne, Australia  
    (e-mail: vera.somers@unimelb.edu.au).}
\address[Second]{Australian Centre for Field Robotics (ACFR), University of Sydney, Australia}

\begin{abstract}          
Spreading processes, e.g. epidemics, wildfires and rumors, are often modeled on static networks. However, their underlying networks structures, e.g changing contacts in social networks, different weather forecasts for wildfires, are due to ever changing circumstances inherently time-varying in nature. In this paper, we therefore, propose an optimization framework for sparse resource allocation for control of spreading processes over temporal networks with known connectivity patterns. We use convex optimization, in particular exponential cone programming, and dynamic programming techniques to bound and minimize the risk of an undetected outbreak by allocating budgeted resources each time step. We demonstrate with misinformation, epidemic and wildfire examples how the method can provide targeted allocation of resources.
\end{abstract}


\end{frontmatter}

\section{Introduction}
Spreading processes have become an influential part of our daily lives. From epidemics (\cite{Nowzari2016}), computer viruses (\cite{Bloem2008} to wildfires (\cite{Karafyllidis1997a}) and misinformation (\cite{zareie2021minimizing}), if not properly monitored and controlled, they can have a disastrous impact and pose a significant threat. This has stressed the importance and current limitation of methods to reduce their risk with appropriate intervention strategies. What these processes or `threats' have in common, is that they can all be modeled as a process with an initial outbreak that spreads rapidly throughout a network.


The natural and social networks these processes typically evolve on, e.g. global contact and travel networks for epidemics, large geographic areas for wildfires, social media or the internet for computer viruses and misinformation, have two important and challenging aspects associated with them. First of all, they are often temporal in nature, i.e. interactions or the connectivity of the network change over time, and second, they evolve on a large scale. This implies it can be difficult or even impossible to deliver resources everywhere quickly across the whole network, such as vaccines for epidemics or waterbombing allocations for wildfires. Therefore, sparse, budgeted and dynamic allocation of resource solutions is important. 

Spreading processes are commonly modeled as Markov processes. The most well-known models are the Susceptible-Infected-Susceptible (SIS) model and the Susceptible-Infected-Removed (SIR) model as proposed by \cite{kermark1927contributions} and \cite{bailey1975mathematical}. These stochastic models can be approximated as ordinary differential equation (ODE) models, which can in turn be approximated by linear models (\cite{Ahn2013}). This is proven by \cite{VanMieghem:2009} to be an upper bound and is, therefore, usually the object of study.  

The problems of minimizing the spreading rate by removing either a fixed number of links or nodes in the network are both NP-hard as discussed in \cite{van2011decreasing}. This could be overcome by heuristics methods, but besides their optimality compromises, the strategy of complete link or node removal is often unrealistic. A more realistic assumption is that spreading rate can be decreased or the recovery rate increased by applying resources to the nodes and links. Various methods have been proposed in which this resource allocation is subject to budget constraints (e.g. see \cite{Nowzari2016,Bloem2008,DiGiamberardino2019,Torres2017,Preciado2014,Zhang2018,Nowzari2017}). 

However, all these methods consider static control of static networks. That is, the resource allocation is determined based on the current state of the network instead of taking into account the time-varying nature of the network. A more realistic approach would be to have time-dependent intervention for temporal networks. Temporarility also implies that the time-varying nature of the spreading variables can be investigated. For example, changing weather conditions for wildfires and varying spreading and recovery rates as different variants emerge or new treatments are developed for epidemics. Furthermore, behavioral changes, growing contact networks or additional external influences can be taken into account by varying the network structure and parameters, which is a step towards a more complete system as indicated in \cite{zino2021analysis}. Furthermore, in \cite{speidel2017epidemic} is stressed that ignoring the temporally-varying nature of networks may underestimate endemicity. 

Dynamic resource allocation for epidemics on static networks is studied in \cite{drakopoulos2017network} and \cite{scaman2016suppressing}. Here network properties are identified to respectively identify limitations on and derive upper and lower bounds for epidemic extinction time. More recently, dynamic programming (\cite{bertsekas2000dynamic}) and in particular, model predictive control (MPC) has been proposed for time dependent intervention (\cite{kohler2018dynamic,watkins2019robust,Selley2015}). 

One method to deal with dynamic networks is by using a time-scale separation. Here the assumption is made that the network dynamics and spreading process evolve on different time-scales, i.e. one process is much faster than the other. In the case of the spreading process being faster, a static network can be used as the changing network structure won't affect the resource allocation and spread. If, on the other hand, the network dynamics are changing faster, we encounter the annealed regime and an `average network' can be used (\cite{zino2021analysis}).

A method that overcomes this time-scale separation assumption are temporal-switching networks. Temporal-switching networks use a sequence of static networks to model the temporal behavior. They were first proposed in \cite{prakash2010virus} for deterministic models, where the epidemic threshold is investigated. In \cite{sanatkar2015epidemic} this is further extended to include the upper bound for epidemic spread and conditions for stability for arbitrarily switching networks. \cite{valdano2015analytical} extend this to continuous time, where the threshold is related to time of observation. 

In \cite{pare2015stability,pare2017epidemic} stability analysis of heterogeneous epidemic processes on time-varying networks is investigated and conditions for global exponential stability and convergence to the disease-free equilibrium are provided. \citet{gracy2020analysis} provide a distributed control strategy by defining conditions for stability and exponential convergence to the disease-free equilibrium. A solution, using geometric programming techniques for control of time-varying networks is given in \cite{Nowzari2016b}.

None of these approaches, however, consider sparse, targeted resource allocation to bound the risk. An exception regarding sparsity being the approach taken in \citet{Rev2R2}, where the authors utilize the reweighted $\ell_1$ optimization approach to solve an antidote problem for a multi-competitive virus model. This approach, however, has no guarantee of global optimality. 

%

In this paper, we therefore, present a framework for dynamic, sparse resource allocation on time-varying networks for control of spreading processes. This framework builds on work presented in \cite{ACC2022}, where we presented a multi-stage sparse resource allocation method for static networks. The presented framework distinguishes itself by extending this work to temporal-switching networks and hence, overcoming the shortcomings of the assumption of a static network. 





\section{Model and Problem Formulation}

\subsection{SIS Spreading Process Model}
We study a spreading process on a time-varying graph $\mathcal{G}(t)$ with $n$ nodes in node set $\mathcal{V}$ and time-variant edge set $\mathcal{E}(t)$. Now, each node $i \in \{1, 2, ..., n\} \in \mathcal{V}$ has a state $X_i(t)$ associated with it. We consider the basic SIS model by \cite{kermark1927contributions} in which a node can be in two states: infected, i.e. $X_i(t)=1$, or susceptible to infection from neighboring nodes, i.e. $X_i(t)=0$. An infected node recovers with probability $\delta_{i}\Delta t$ to $X_i(t+\Delta t)=0$ and the process spreads from infected node $j$ to susceptible node $i$ with probability $\beta_{ij} \Delta t$, where $\Delta t$ is a small time interval. Due to the time-varying nature of our spreading process and network, we extend this further by allowing $\beta_{ij}(t)$ and $\delta_i(t)$ to change over time. 

We now define $x_{i}(t)=E(X_{i}(t))=P(X_{i}(t)=1)$ as the probability of a node $i$ being infected at time $t$. Following \cite{Preciado2014}, with a mean-field approximation, $n$ coupled nonlinear differential equations are obtained:
\begin{equation}
\label{eq:nonL}
\dot{x}_{i}=(1-x_{i}(t))\sum^{n}_{j=1}\beta_{ij}(t)x_{j}(t)-\delta_{i}(t)x_{i}(t).
\end{equation}
By applying Euler's forward approximation to \eqref{eq:nonL}, see \cite{pare2020modeling}, we obtain the following discrete time model approximation 
\begin{equation}
\label{eq:disnonL}
x_{i}^{k+1}=x_i^k + h(1-x_{i}^k)\sum^{n}_{j=1}\beta^k_{ij}x_{j}^k-h\delta^k_{i}x_{i}^k
\end{equation}
where $h= t^{k+1}-t^k$ is the length of each time interval $[t^k, t^{k+1}]$. Given $h\delta^k \leq 1$ and $h\sum^{n}_{j=1}\beta^k_{ij}<1 \,\, \forall\, k$, this model is well-defined at $x_i^1\in[0,1] \Rightarrow x_i^k\in[0,1] \,\, \forall\, k$.


By linearizing around the infection-free equilibrium point ($x_i=0 \,\, \forall\, i$) we obtain
\begin{equation}
\label{eq:disL}
x^{k+1}=A^kx^k
\end{equation}
where $x^k=[x_{1}^k,...,x_{n}^k]^{T}$ is the state of the system and $A^k$ the sparse state matrix consisting of elements
 \begin{equation}
  \label{eq:epi}
 a_{ij}^k = 
\begin{cases}
1-h\delta_{i}^k  \ge 0 &\quad \text{if}\quad  i=j, \\
h\beta_{ij}^k \ge 0&\quad \text{if}\quad  i\neq j, (i,j) \in \mathcal{E}^k, \\
0  &\quad \text{otherwise}.
 \end {cases}
 \end{equation}
Note that the state matrix $A^k$ changes each time step due to both changes in spreading parameters $\beta^k_{ij}$, $\delta_i^k$ and due to changes in network structure, i.e. in $\mathcal{E}^k$. Hence, we have obtained a system that can take a sequence of temporal-switching graphs $\mathcal{G}^k(\mathcal{V},\mathcal{E}^k)$ as input.
 
Furthermore, we remark that \eqref{eq:disL} is a positive system as described in \cite{berman1994nonnegative}, i.e. if $x^1_i\ge 0$ for all $i$, then $x^i_k\ge 0$ for all $i$ and all $k\ge 0$.

%
  
\subsection{Cost function}
The aim is to reduce and bound the risk of the spreading system on the temporal-graph described in the previous section. In order to achieve this, we assume a finite-time horizon forecast and optimization up to time $K$ and associate the following cost function with the system  
\begin{equation}
\label{eq:JC}
  J(x^1) = \sum_{k=1}^K \alpha^kCx^k
\end{equation}
where $x^k=[x_{1}^k,...,x_{n}^k]^{T}$ is the state of the system and $C= [c_1, ..., c_n]$ is a row vector defining the cost at time $k$ associated with each node $i$, with each $c_i\ge 0$. We include a discount factor $\alpha\in(0,1]$ which can be tuned to emphasize near-term cost over long-term cost. Note that the cost for $K \rightarrow \infty$ can also be taken into account as further explained in Section \ref{subsec:Inf}.

We can now formulate bounds on the cost \eqref{eq:JC} based on the linear model \eqref{eq:disL} and demonstrate it also bounds the costs of the nonlinear model \eqref{eq:disnonL}. Positive systems with non-negative linear costs admit linear value, or cost-to-go, functions, see e.g. \cite{berman1994nonnegative, briat2013robust, rantzer2015scalable}, and we will use a variation of this argument to construct bounds via a dynamic-programming-like formulation. %

\begin{prop}\label{prop1}
	Suppose there exists a sequence of row vectors $p^k=[p_{1}^k,...,p_{n}^k], k = 1, ..., K$ which are elementwise non-negative: $p_i^k\ge 0 \,\, \forall\, i,k$, and for which the following inequality holds:
\begin{equation}
\label{eq:pk1}
  p^k \ge C+\alpha p^{k+1}A^{k}
\end{equation}
where the inequalities are understood to hold elementwise. Then $p^1x^1\ge J(x^1)$ as defined in \eqref{eq:JC}.
\end{prop}
\begin{pf}
We first show that $V(x,k) = p^kx^k$ provides an upper bound on the cost-to-go from state $x$ at time $k$ for the linear system \eqref{eq:disL}. Since $x_i^k\ge 0 \,\, \forall\, i,k$, inequality \eqref{eq:pk1} implies
$
  p^kx^k \ge Cx^k+\alpha p^{k+1}A^{k}x^k.
$
Using $V(x,k) = p^kx^k$, we obtain
\begin{equation}
\label{eq:vk1}
  V(x^k,k) \ge Cx^k+\alpha V(x^{k+1},k+1).
\end{equation}
By telescoping sum we obtain
$
  V(x^1,1) \ge \sum_{k=1}^K \alpha^kCx^k + V(x^K,K+1)
$
 and since $V(x^K,K+1)\ge 0$ we have
$
  V(x^1,1) = p^1x^1 \ge \sum_{k=1}^K \alpha^kCx^k.
$

To demonstrate this also bounds the cost for the nonlinear system \eqref{eq:disnonL}, we note that for any $x^k$ with elements in $[0,1]$, $x^{k+1}\ge A^kx^k$ where $x^{k+1}$ is evaluated according to \eqref{eq:disnonL}, i.e. solutions of the linear system upper bound (elementwise) those of the nonlinear system from the same initial conditions. Because the cost vector $C$ is positive, this further implies that the cost obtained with the linear system upper bounds that with the nonlinear system.
\end{pf}

To go from these cost bounds to a model of risk, we note that we can bound the expected cost $E[J(x^1)]$ via a linear function of the expected value of the initial state $\hat x^1$. Because we have $J(x^1) \le p^1x^1$ for any $x^1$, it follows that $E[J(x^1)] \le E[p^1x^1] = p^1 E[x^1]=p^1\hat x^1$. We, therefore, define the risk associated with each state as
\begin{equation}
\label{eq:Risk2}
   R_i=p_i^1\hat{x}_i^1.
\end{equation}
Both the overall risk $\sum_i R_i=p^1\hat x^1$ and the maximum risk or `worst-case' expected impact of an outbreak $\max_{i}(R_{i})$ can be considered.


\subsection{Resource Allocation Model}
In order to reduce the overall cost and risk involved, we now investigate the inclusion of sparse, budgeted resources at each time step $k$ to our model. We consider allocation of resources to both the edges, to reduce spreading rates $\beta_{ij}$, and to the nodes, i.e. to increase recovery rates $\delta_{i}$. In order to achieve this, we assume bounded ranges of possible spreading and recovery rates for $k=1,...,K$:
\begin{equation}
 \label{eq:bounds2}
0 < \underline{\beta}_{ij}^k \leq \beta_{ij}^k \leq \overline{\beta}_{ij}^k, \quad 0 < \underline{\delta}_{i}^k \leq \delta_{i}^k \leq \overline{\delta}_{i}^k<\overline{\Delta}.
\end{equation}
Here $\overline{\beta}_{ij}^1$ and $ \underline{\delta}_{i}^1 $ are the unmodified rates of the system \eqref{eq:disnonL}.  For $k>1$,  $\overline{\beta}_{ij}^k$ and $ \underline{\delta}_{i}^k $ are the updated rates of the system given the unmodified rates associated with time step $k$ with the inclusion of resource allocation up to the current time step. That is, in our model, changes made at a particular time step persist for all future times. This implies that our state matrix $A^k$ now changes each time step not only due to the time-varying nature of the graph as set out in \eqref{eq:epi}, but also due to inclusion of resource allocation on $\beta_{ij}$ and $\delta_i$ up to time $k$. Note that the framework can easily be modified to include interventions with a fixed duration or those that diminish over time.

We now propose the following resource model defined by:
\begin{align}
f_{ij}\left(\beta_{ij}^k\right)&=w_{ij}^k\text{log}\left(\frac{\overline{\beta}^k_{ij}}{\beta_{ij}^k}\right) \nonumber \\
g_{i}\left(\overline{\Delta}-\delta_{i}^k\right)&=w_{ii}^k\text{log}\left(\frac{\overline{\Delta}-\underline{\delta}^k_{i}}{\overline{\Delta}-\delta_{i}^k}\right) \label{eq:RM}
\end{align}
where $w_{ij}$ are weightings expressing the cost of respectively reducing $\beta_{ij}$ and increasing $\delta_{i}$ at time $k$.

These resource models can be understood as that a certain proportional increase (for $\beta_{ij}$) and decrease (for $\delta_i$) always have the same cost. Note that it is impossible for $\beta_{ij}$ to be reduced to $0$ since that would take infinite resources.  More importantly, these resource models encourage sparsity as detailed further in Section \ref{subsec:spars}. We refer to \cite{LCSS2021} for a more in-depth discussion of this type of logarithmic resource allocation.
  
\subsection{Problem Statements}
The goal is to bound both the risk of an outbreak and the amount of resources required while adhering to the time-varying network structure. Therefore, we study two closely related problems of dynamic resource allocation for spreading processes on temporal-switching graphs. 

\begin{prob}[Resource-Constrained Risk Minimization] \label{prob:1}
\quad \\
Given a sequence of temporal-switching graphs $\mathcal{G}^k(\mathcal{V},\mathcal{E}^k)$, a defined resource allocation budget $\Gamma^k$ per time step $k$, a total budget $\Gamma_{\text{tot}}$ and a cost $c_{i}$ associated with each node $i \in \mathcal{V}$, find the optimal spreading and recovery rates $\beta_{ij}^k$ and $\delta_{i}^k$ for $k=1, ..., K$ that via sparse resource allocation minimize an upper bound on the maximum risk $p_i^1\hat{x}_i^1$. This problem can be formulated as:
\begin{align}
    \underset{p^k,\beta^k, \delta^k}{\text{minimize}} & \quad   \text{max}(p_i^1\hat{x}_i^1) \label{eq:C0}\\
    \text{such that} & \quad p^k \geq 0, \quad \alpha p^{k+1} A^k - p^{k} \leq -C, \label{eq:C1a}\\
    & \quad 0 < \underline{\beta}_{ij}^k \leq \beta_{ij}^k \leq \overline{\beta}_{ij}^k, \ 0 < \underline{\delta}_{i}^k \leq \delta_{i}^k \leq \overline{\delta}_{i}^k < \overline{\Delta}, \label{eq:rCons} \\
 & \quad  \sum_{ij} f_{ij}\left(\beta_{ij}^k\right) + \sum_{i} g_{i}\left(\delta_{i}^k\right) \leq \Gamma^k, \label{eq:C2} \\
 & \quad \sum^K_{k=1} \biggl(\sum_{ij} f_{ij}\left(\beta_{ij}^k\right) + \sum_{i} g_{i}\left(\delta_{i}^k\right)\biggr) \leq \Gamma_{\text{tot}}.
\end{align} 
\end{prob}

\begin{prob}[Risk-Constrained Resource Minimization] \label{prob:2}
\quad \\
Find the optimal spreading and recovery rates $\beta_{ij}^k$ and $\delta_{i}^k$ for $k=1, ..., K$ that via sparse resource allocation minimize the amount of resources required, given an upper bound on the maximum risk $\gamma$, a sequence of temporal-switching graphs $\mathcal{G}^k(\mathcal{V},\mathcal{E}^k)$ and a cost $c_{i}$ for $i \in \mathcal{V}$. 

This problem can be formulated similarly to optimization Problem \ref{prob:1} except with the left-hand-side of \eqref{eq:C2} as an objective and a constraint $p_i^1\hat{x}_i^1 \leq \gamma$ for all $i$.
\end{prob}

This problem formulation can be extended by including surveillance scheduling and resource allocation on $\hat{x}_i^1$, see \cite{TCNS2021} for details. Note also that time $K \rightarrow \infty$ can be included, see Section \ref{subsec:Inf}, and we can replace \eqref{eq:C0} with $\sum_ip_i^1\hat{x}_i^1$ to bound overall risk.
 
\section{Exponential Cone Programming}
In this section we show that Problems \ref{prob:1} and \ref{prob:2} can be reformulated as convex optimization problems, in particular exponential cone programs. Furthermore, we discuss how the proposed resource model leads to sparse resource allocation. 

We define decision variables $y_i$ and $v_i$ for all nodes $i\in \mathcal{V}$, and $u_{ij}^k$ for all edges $(i,j)\in\mathcal E^k$. 

\begin{prop} 
\label{prop:1}
Given that $h\overline{\Delta} < 1$, Problem \ref{prob:1} is equivalent to the following convex optimization problem under the transformations $y_i=\text{log}(p_i)$ and $u_{ij}=f_{ij}\left(\beta_{ij}\right)$ and $v_{i}=g_{i}\left(\overline{\Delta} - \delta_{i}\right)$
\begin{align}
     \underset{y^k, u^k, v^k}{\text{minimize}} & \quad  \text{max}(\text{log}(\hat{x}_i^1)+ y_i^1) \label{eq:OF1}\\
    \text{such that} & \quad \text{log} \Biggl (\sum_{i: (i,j)\in \mathcal E} \text{exp} \Biggl(y_{i}^{k+1} + \text{log}\left(\alpha h\overline{\beta}^k_{ij}\right) - \frac{u_{ij}^k}{w_{ij}}   \nonumber\\ 
& \quad -  y_{j}^k \Biggr) + \text{exp}\left(y_{j}^{k+1} + \text{log}\left(\alpha(1-h\overline{\Delta})\right) - y_{j}^k\right)  \nonumber \\
&  \quad + \text{exp}\left(y_{j}^{k+1}+ \text{log}\left(\alpha h(\overline{\Delta}-\underline{\delta})\right) - \frac{v_{i}^k}{w_{ii}} - y_{j}^k\right) \nonumber \\
& \quad + \text{exp}\left (\text{log}\left (c_{j}\right) - y_{j}^k\right) \Biggr) \leq 0 \quad \forall j,   \label{eq:CNew} \\
& \quad 0 \leq u_{ij}^k \leq w_{ij}\text{log}\left (\frac{\overline{\beta}_{ij}^k}{ \underline{\beta}_{ij}^k}\right ), \label{eq:CB} \\
& \quad 0 \leq v_{i}^k \leq w_{ii}\text{log}\left (\frac{1-\underline{\delta}^k_{i}}{ 1-\overline{\delta}^k_{i}}\right ) ,\label{eq:CD} \\
& \quad  \sum_{ij} u_{ij}^k + \sum_{i} v_{i}^k \leq \Gamma^k, \label{eq:C5} \\
& \quad \sum^K_{k=1} \biggl(\sum_{ij} u_{ij}^k + \sum_{i} v_{i}^k\biggr) \leq \Gamma_{\text{tot}} \label{eq:C6}.
\end{align}
\end{prop}

\begin{pf}
Similar to the proof of Proposition 2 in \cite{ACC2022}.
\end{pf}

\begin{prop} 
Given that $h\overline{\Delta} < 1$, Problem \ref{prob:2} is equivalent to a convex optimization problem under the transformations $y_i=\text{log}(p_i)$ and $u_{ij}=f_{ij}\left(\beta_{ij}\right)$ and $v_{i}=g_{i}\left(\overline{\Delta}- \delta_{i}\right)$. The formulation is the same as \eqref{eq:OF1} - \eqref{eq:C6} except that the objective is $\sum_{ij} u_{ij}^k + \sum_{i} v_{i}^k$ and in place of \eqref{eq:C5} we have the constraint $ \text{log}(\hat{x}_i^1)+ y_i^1 \leq \gamma$.
\end{prop}

\begin{pf}
See \cite{ACC2022}.
\end{pf}

\subsection{Sparsity}
\label{subsec:spars}
A significant benefit of the proposed framework and exponential cone programming formulation is that it encourages sparse resource allocation. Our resource models, now formulated as constraints \eqref{eq:C5} and \eqref{eq:C6} are $\ell_{1}$ norm constraints and objectives, since $u_{ij}\geq 0$ and $v_i \geq 0$, which are known to encourage sparsity (e.g. \cite{Candes2008} and references therein).

If the goal is maximal sparsity, i.e. minimal number of nodes and edges with non-zero resources allocated, then the reweighted $\ell_{1}$ optimization approach of \cite{Candes2008} can be used. We can include this in our model by iteratively solving our problem, but with a reweighted resource model that approximates the number of nodes and edges with non-zero allocation, e.g.
\begin{equation}
\label{eq:L1} 
\phi^q=\sum_{ij} \frac{u^{k^q}_{ij}}{u_{ij}^{k^{q-1}}+\epsilon}+\sum_{i} \frac{v^{k^q}_{i}}{v_{i}^{k^{q-1}}+\epsilon}
\end{equation} 
where $q$ is the iteration number and $\epsilon$ a small positive constant to improve numerical stability. For Problem \ref{prob:1} we now replace \eqref{eq:C5} with $\phi^q \leq M$ where $M$ is the bound on the number of nodes and edges that can have resources allocated to them. Similarly for Problem \ref{prob:2} we replace the objective with \eqref{eq:L1}. This iterative approach can ignore previous budget bounds and has no guarantee of convergence or global optimality, but has been found to be effective in practice. 
%
%

\subsection{Infinite Time} 
\label{subsec:Inf}
The proposed framework considers the scenario of a finite-time horizon optimization up to time step $K$, which makes sense given practical examples often have limited time (reliable) forecasts. Alternatively, the framework can easily be extended to include the cost as $K \rightarrow \infty$. The updated cost bounds (see Proposition \ref{prop:1}) can be found as follows.

\begin{prop}
Given an upperbound on future network changes, i.e. on future values of $A$, $A^K \geq A^k$ for all $k \geq K$, and suppose there exists a sequence of row vectors $p^k=[p_{1}^k,...,p_{n}^k], k = 1, ..., K$ which are elementwise non-negative: $p_i^k\ge 0 \,\, \forall\, i,k$, and for which inequality \eqref{eq:pk1} now holds for $k=1,...,K-1$ and 
\begin{equation}
\label{eq:pK1}
  p^K \ge C+\alpha p^K A^{K} \geq C + \alpha p^K A^k,
\end{equation}
where the inequalities are understood to hold elementwise. Then $p^1x^1\ge J(x^1)$ as defined in \eqref{eq:JC}.
\end{prop}

\begin{pf}
The second inequality in \eqref{eq:pK1} follows because $p$ is non-negative and $A^K \geq A^k$. In the proof of Proposition \ref{prop:1} we obtained \eqref{eq:vk1} from \eqref{eq:pk1} and similarly \eqref{eq:pK1} implies
\begin{equation}
  V(x^k,K) \ge Cx^k+\alpha V(x^{k+1},K)
\end{equation}
for $k \ge K$. By telescoping sum we now obtain
$
  V(x^1,1) \ge \sum_{k=1}^L \alpha^kCx^k + V(x^L,K)
$
for any $L\ge K$, and since $V(x^L,K)\ge 0$ and $L$ is arbitrarily large we have
$
  V(x^1,1) = p^1x^1 \ge \sum_{k=1}^\infty \alpha^kCx^k.
$
\end{pf}

Our optimization Problems \ref{prob:1} and \ref{prob:2} now use \eqref{eq:C1a} for $k=1,...,K-1$ and include the additional constraint
\begin{equation}
p^{K} (\alpha A^K-I) \le -C.
\label{eq:DCK}
\end{equation}
For the corresponding convex formulation we refer to equation (21) in \cite{ACC2022}.


%
%

\section{Results}
In this section we demonstrate how our proposed optimization framework can be utilized to investigate resource allocation on temporal-switching networks. First, we discuss a 7-node example to illustrate the differences in resource allocation for assuming a static, average and temporal-switching network. Next, we show a periodic flight network and corresponding sparse flight reduction to reduce the risk of a known SARS outbreak. Finally, we present a wildfire example with changing wind forecast.

\subsection{Small Influence Network}
\begin{figure}[!t]
\centering
 \def\svgwidth{0.7\linewidth}
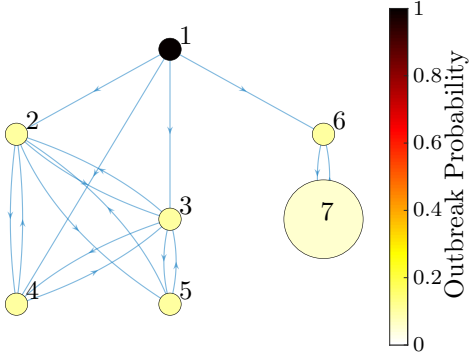   
\caption{Influence graph with $n=7$ nodes. Node color indicates outbreak probability $\hat{x}_i^1$, node marker size indicates cost $c_i$ with high cost node $i=7$.}
\label{fig:GraphAM1}
\end{figure}

We, first, start with a small 7-node example as it is easier to visualize network sequence changes and impact on a small scale. In this particular example, we model the spread of misinformation and consider the influence graph $\mathcal{G}^k=(\mathcal{V},\mathcal{E}^k)$ with $n=7$ nodes as visualized for $k \geq K$ in Fig. \ref{fig:GraphAM1}. Here node $i=1$ has a high probability $\hat{x}_i^1$, indicated by node color, of starting this particular rumor and arrows indicate the direction of influence, e.g. node $i=1$ influences node 6, but not the other way around, while node 6 and 7 share (mis)information with each other. The actual influence network, however, changed over time $k\leq K$, with nodes making new connections or e.g. stopping to follow other nodes due to unwanted content. We take homogeneous spreading and recovery rates of $\beta_{ij}^k=0.35$ for all $(i,j) \in \mathcal{E}^k$ and $\delta_i=0.2$ for all $i \in \mathcal{V}$. Furthermore, we introduce a high cost node $i=7$ with $c_i=1$, as indicated by node marker size, where we don't want this rumor to end up due to potentially their connections to other larger media networks or vulnerability to misinformation (e.g. scams or anti-vaccine campaigns to elderly). 


The goal is now to both stop the spread of this rumor and in particular to protect the high cost node $i=7$. The impact of rumors and misinformation can be reduced by `truth campaigns', i.e. increasing awareness or recovery rate $\delta$, and by blocking incoming `risky' content, i.e. reducing the incoming spreading rate $\beta$ (\cite{zareie2021minimizing}). We model this simultaneously where targeted nodes have both increased recovery rate and reduced spreading rate. Setting $\overline{\Delta}=1$, $\alpha=0.93$, $h=0.24$ and $\Gamma^k=1.5$ for all $k$ with $K=4$, we first solve Problem 2. Next, to emphasize sparsity and difference in outcomes, we use the reweighted $\ell_1$ minimization approach, see Section \ref{subsec:spars}, by keeping the risk bound but minimizing the number of non-zero allocations. 

\begin{figure}
\centering
\begin{subfigure}[b]{0.45\linewidth}
 \def\svgwidth{1\linewidth}
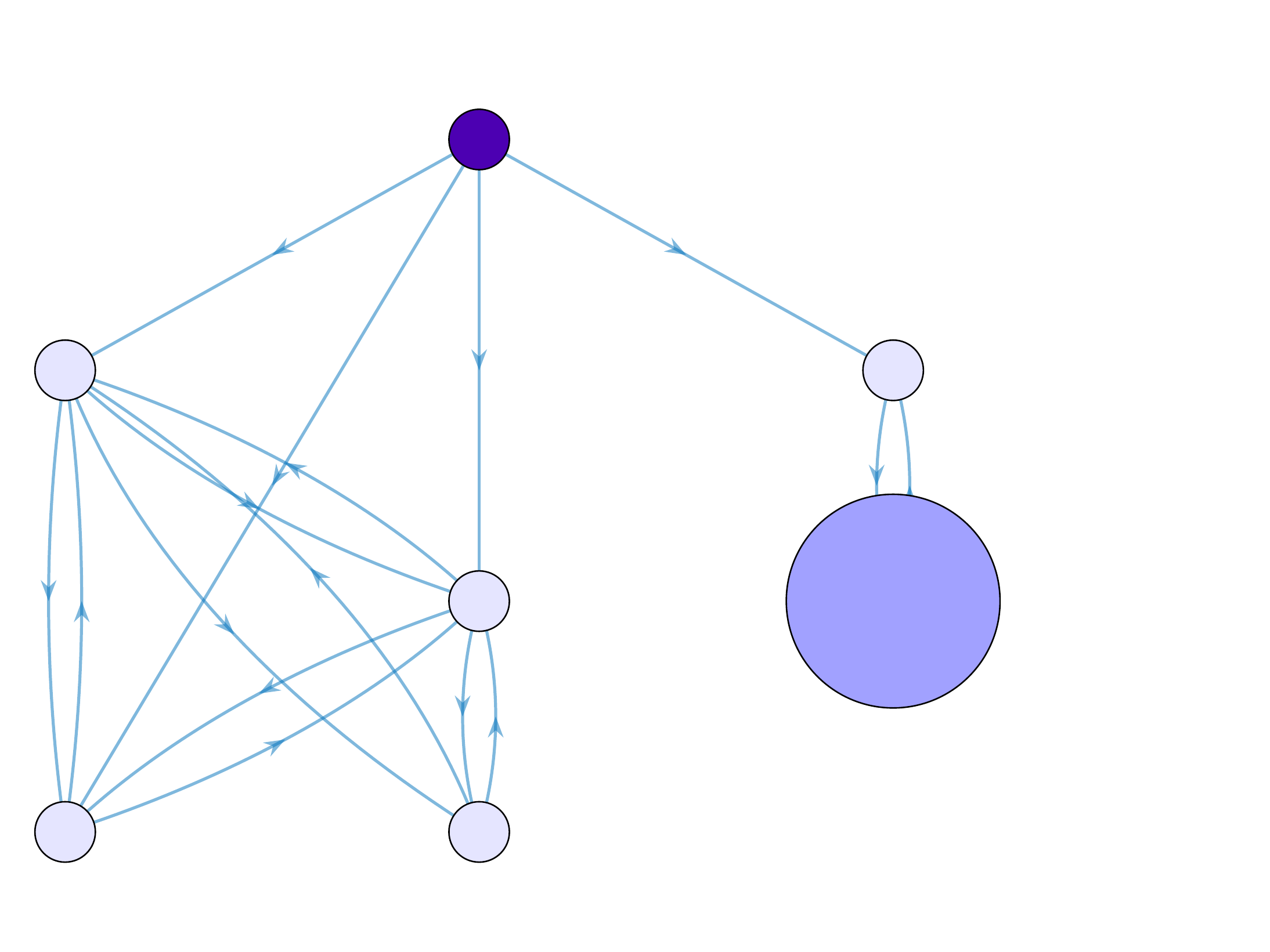
      \caption{k=1}
      \label{fig:031}
  \end{subfigure}
  ~ 
     \begin{subfigure}[b]{0.45\linewidth}
 \def\svgwidth{1\linewidth}
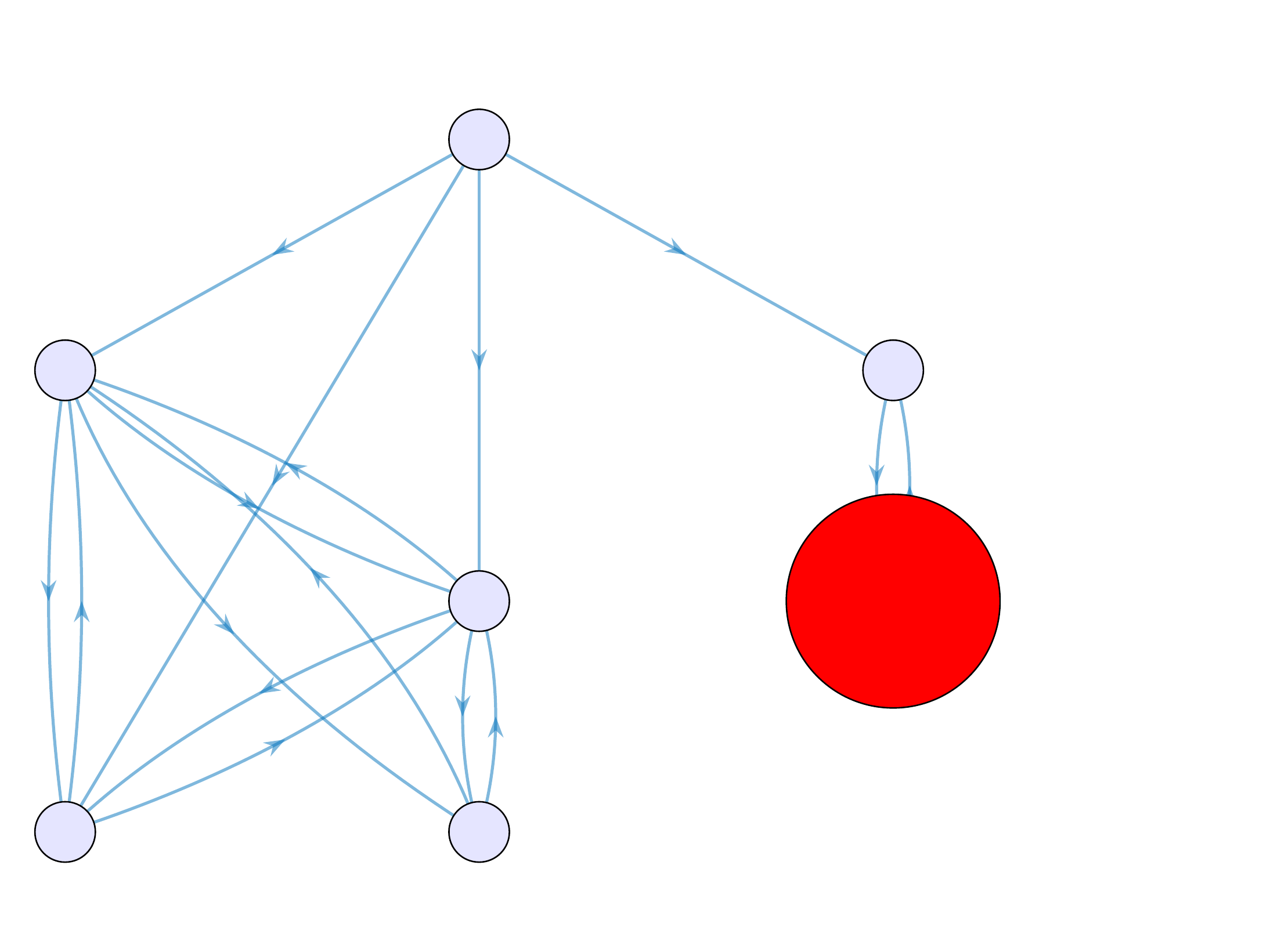
\caption{k=2}
      \label{fig:032}
  \end{subfigure}
  \begin{subfigure}[b]{0.45\linewidth}
 \def\svgwidth{1\linewidth}
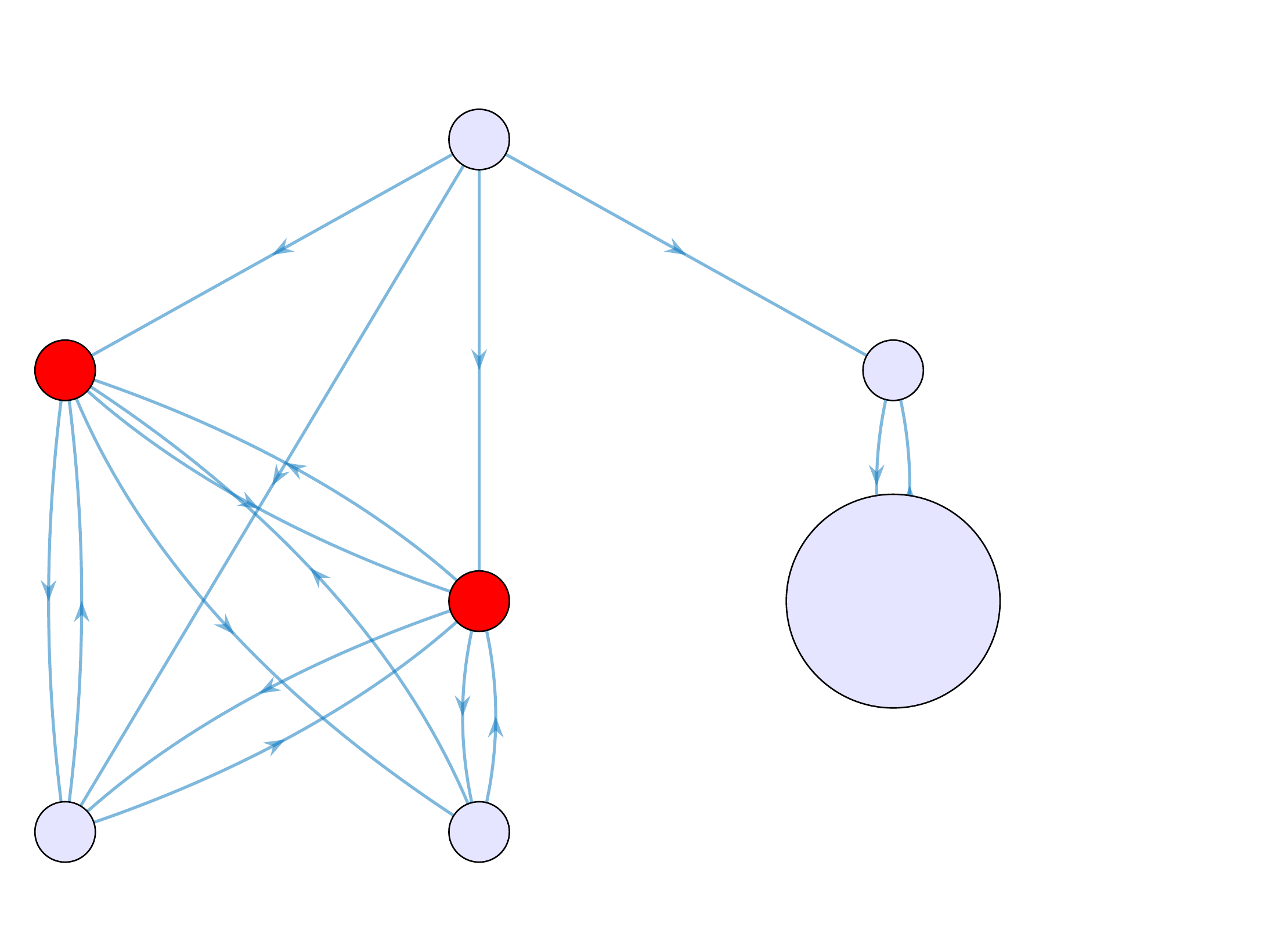
      \caption{k=3}
      \label{fig:033}
  \end{subfigure}
  ~ 
     \begin{subfigure}[b]{0.45\linewidth}
 \def\svgwidth{1\linewidth}
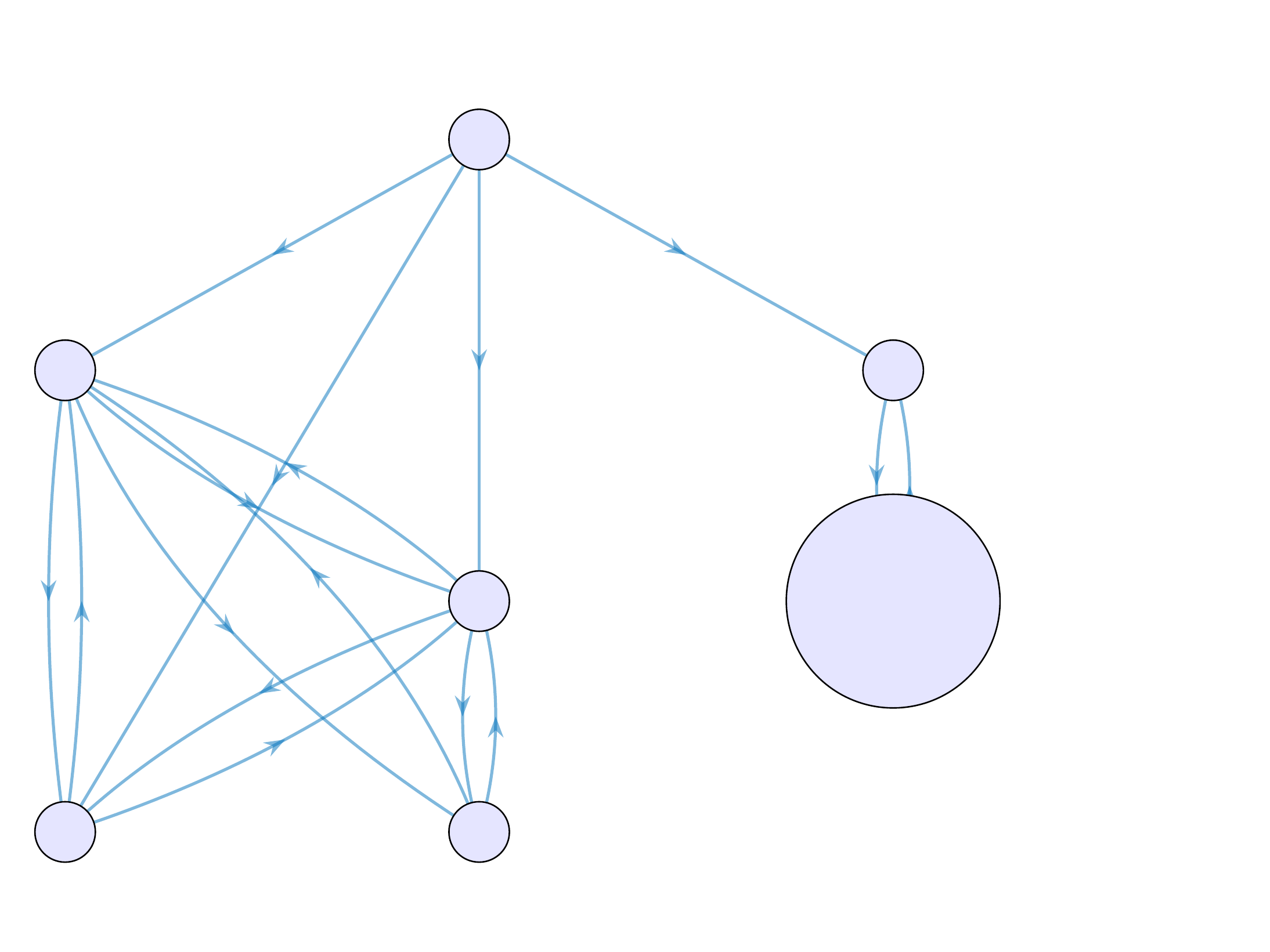
\caption{k=K}
      \label{fig:034}
  \end{subfigure}
      \caption{Anti-rumor strategy using a \emph{static} network with $h=0.24$, $\alpha=0.93$. Node color indicates nodes targeted.}     
\label{fig:03}
\end{figure}
\begin{figure}
\centering
\begin{subfigure}[b]{0.45\linewidth}
 \def\svgwidth{1\linewidth}
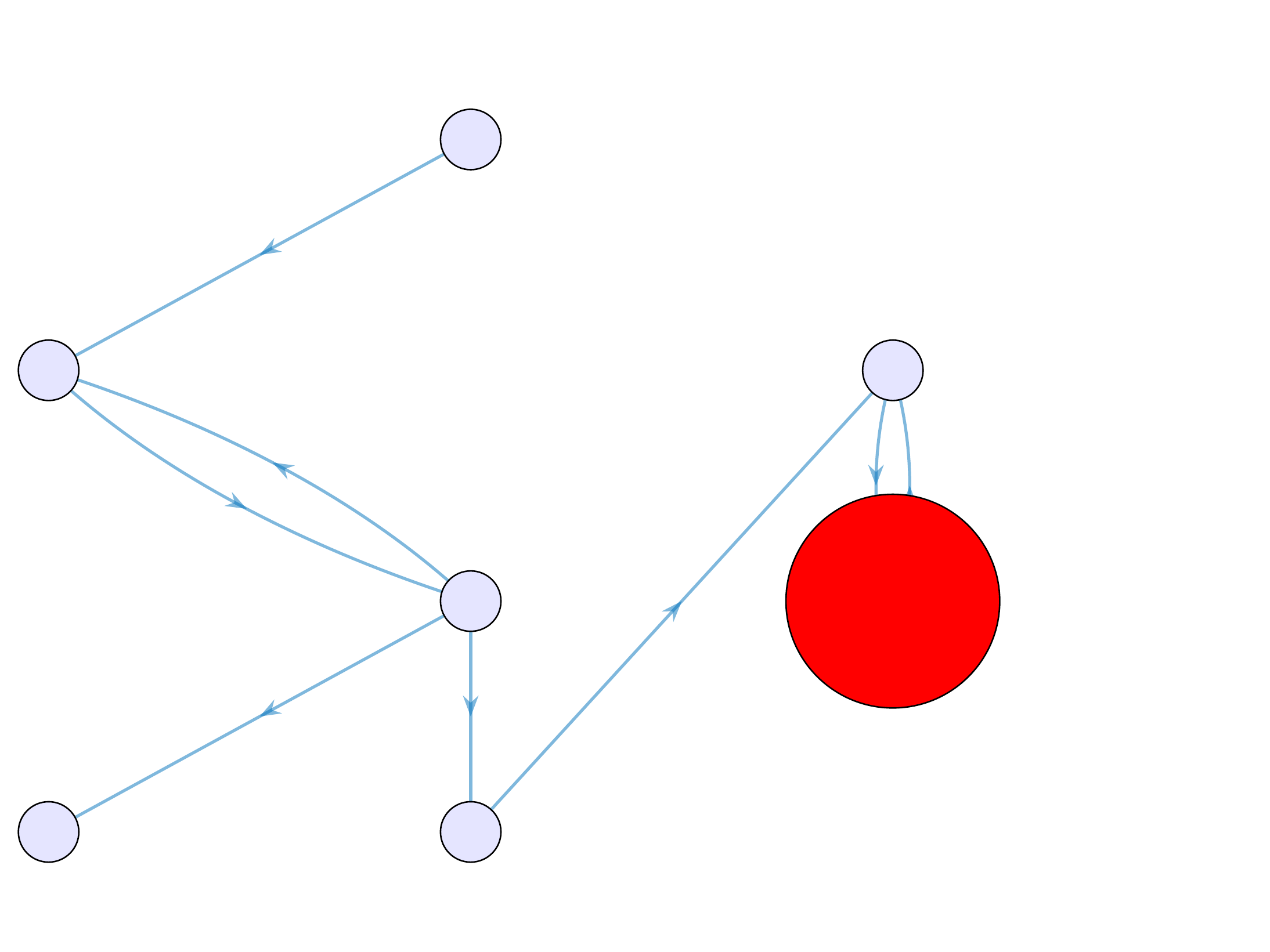
      \caption{k=1}
      \label{fig:031}
  \end{subfigure}
  ~ 
     \begin{subfigure}[b]{0.45\linewidth}
 \def\svgwidth{1\linewidth}
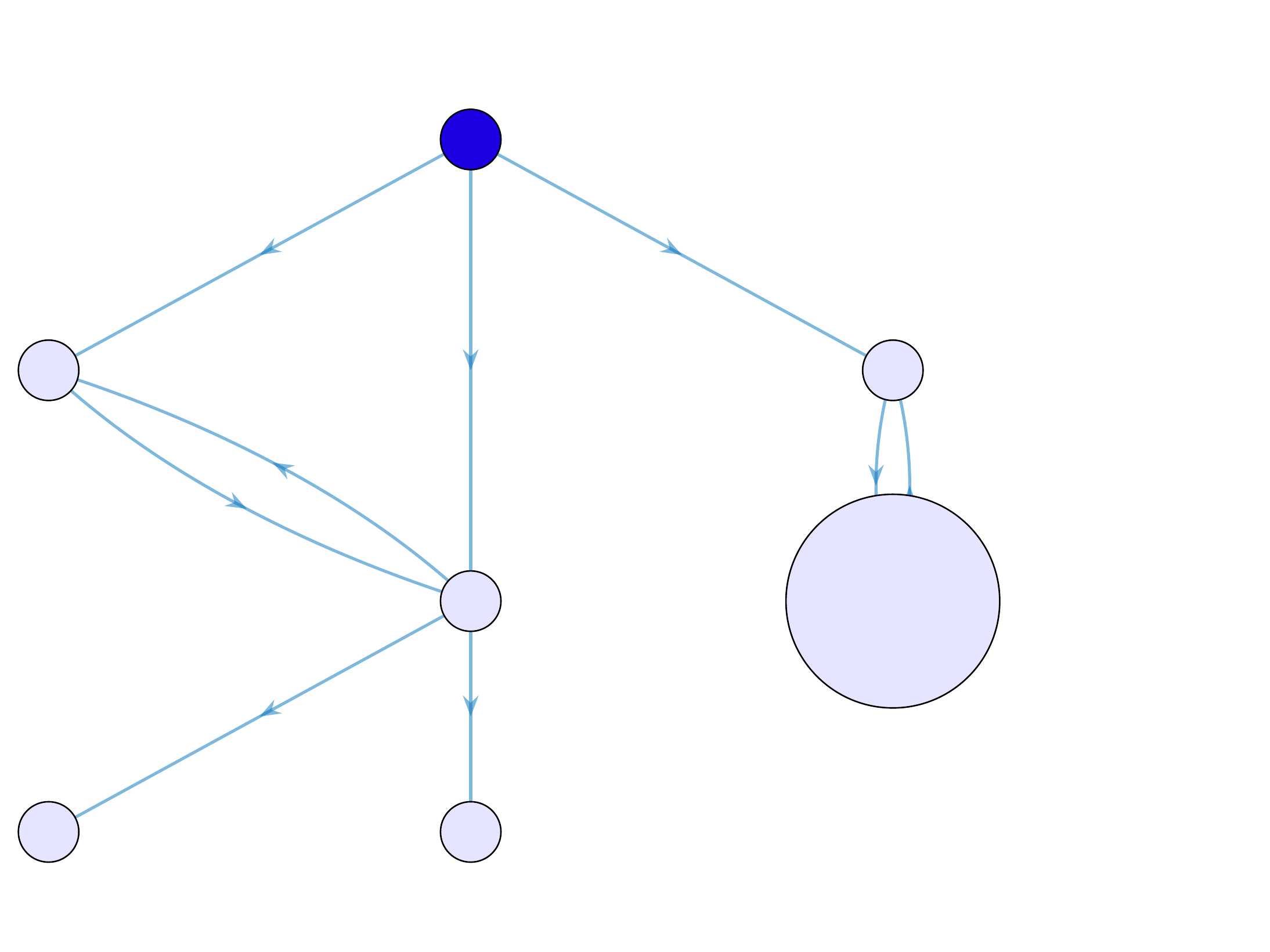
\caption{k=2}
      \label{fig:032}
  \end{subfigure}
  \begin{subfigure}[b]{0.45\linewidth}
 \def\svgwidth{1\linewidth}
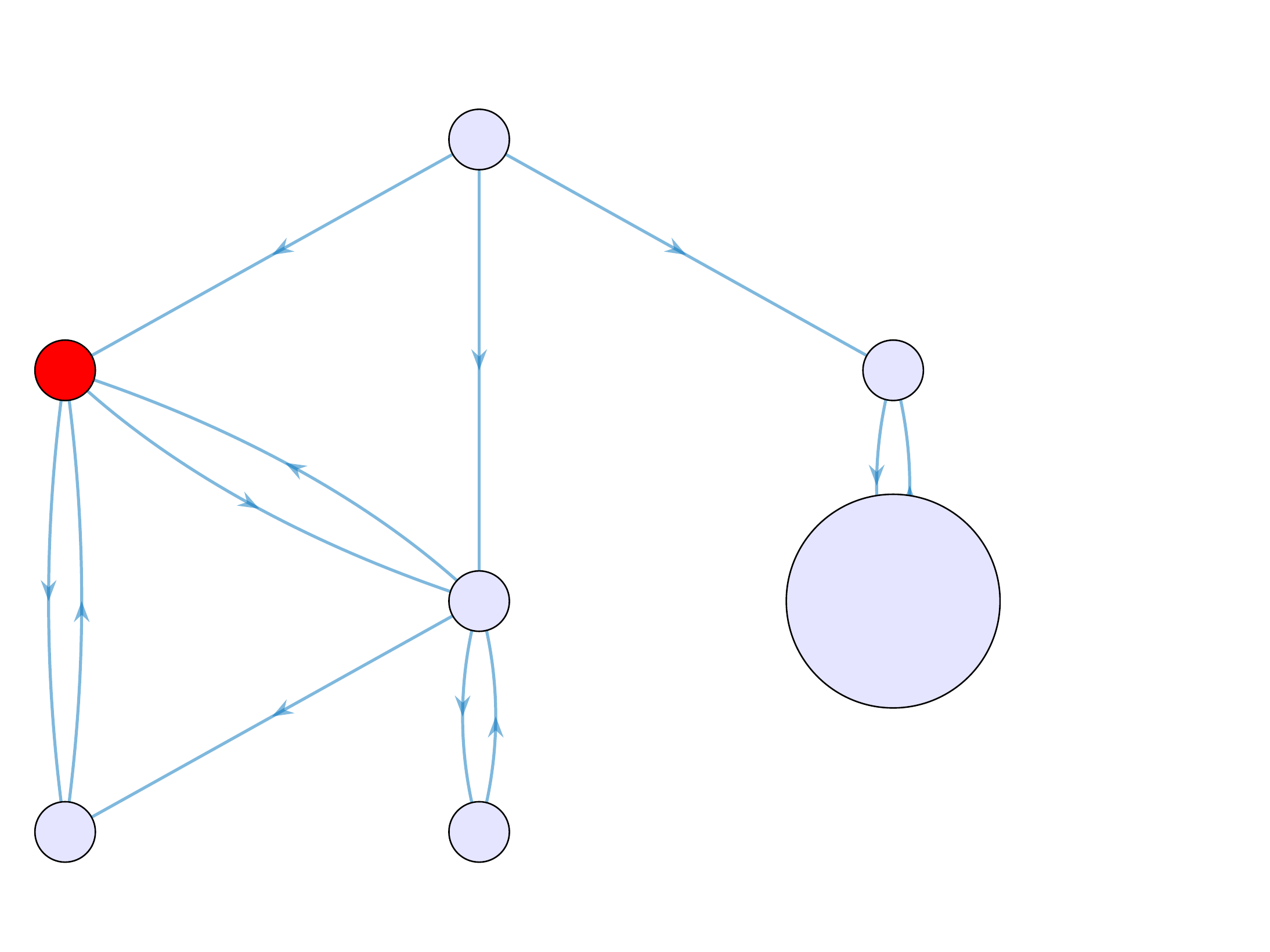
      \caption{k=3}
      \label{fig:033}
  \end{subfigure}
  ~ 
     \begin{subfigure}[b]{0.45\linewidth}
 \def\svgwidth{1\linewidth}
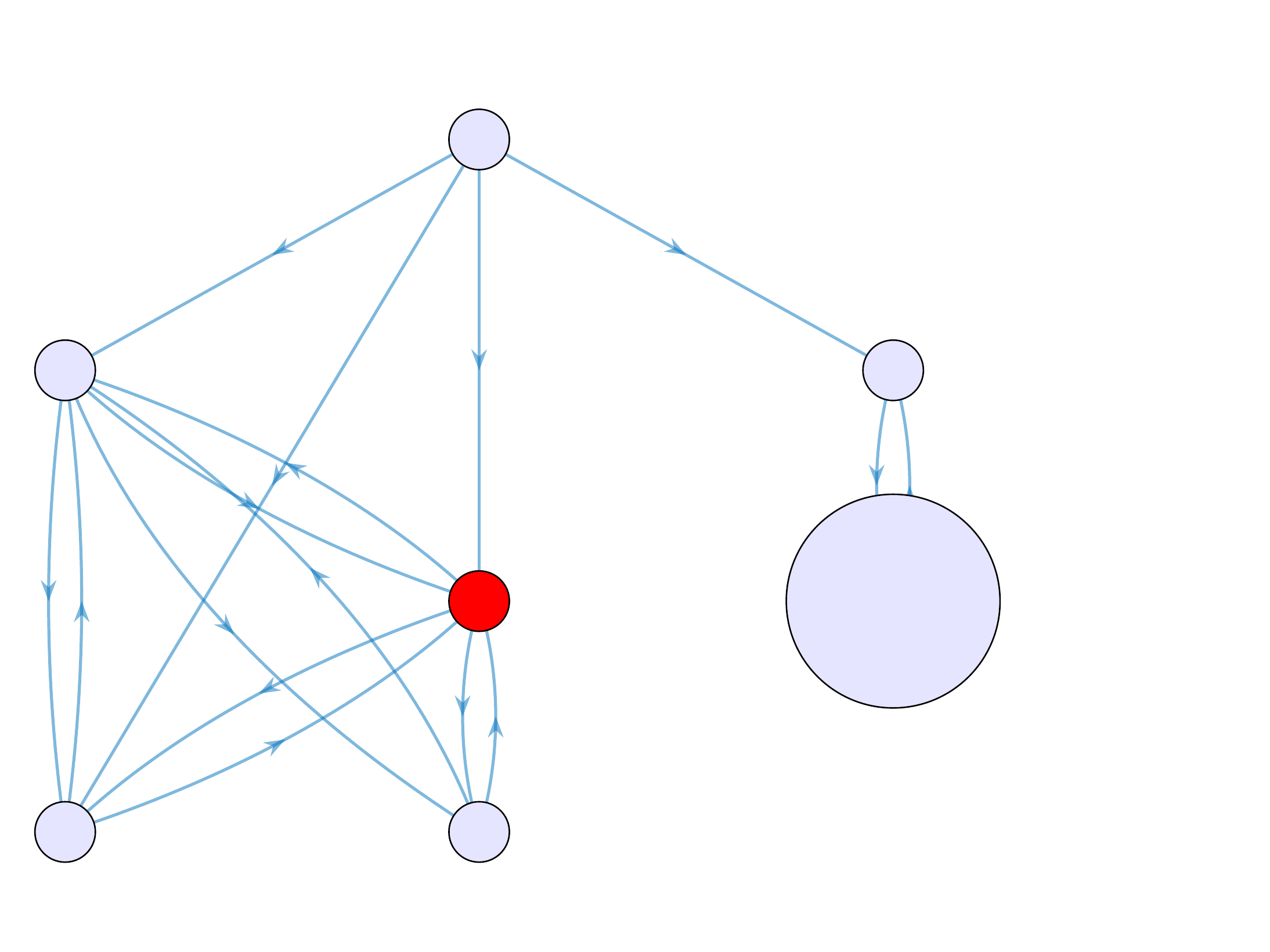
\caption{k=K}
      \label{fig:034}
  \end{subfigure}
      \caption{Anti-rumor strategy for a \emph{temporal-switching} network sequence with $h=0.24$, $\alpha=0.93$.}     
\label{fig:KRA}
\end{figure}
\begin{figure}
\centering
\begin{subfigure}[b]{0.45\linewidth}
 \def\svgwidth{1\linewidth}
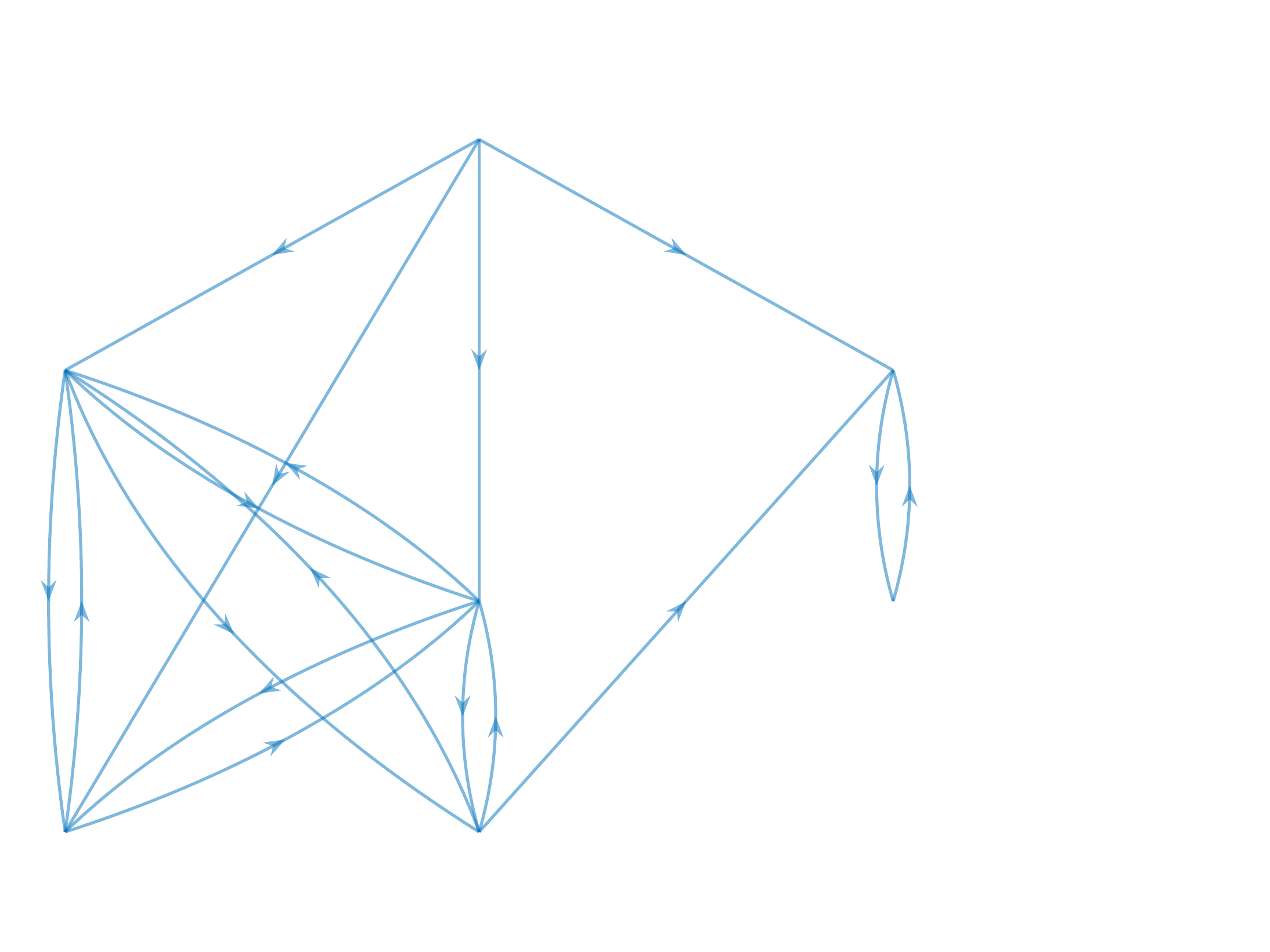
      \caption{k=1}
      \label{fig:031}
  \end{subfigure}
  ~ 
     \begin{subfigure}[b]{0.45\linewidth}
 \def\svgwidth{1\linewidth}
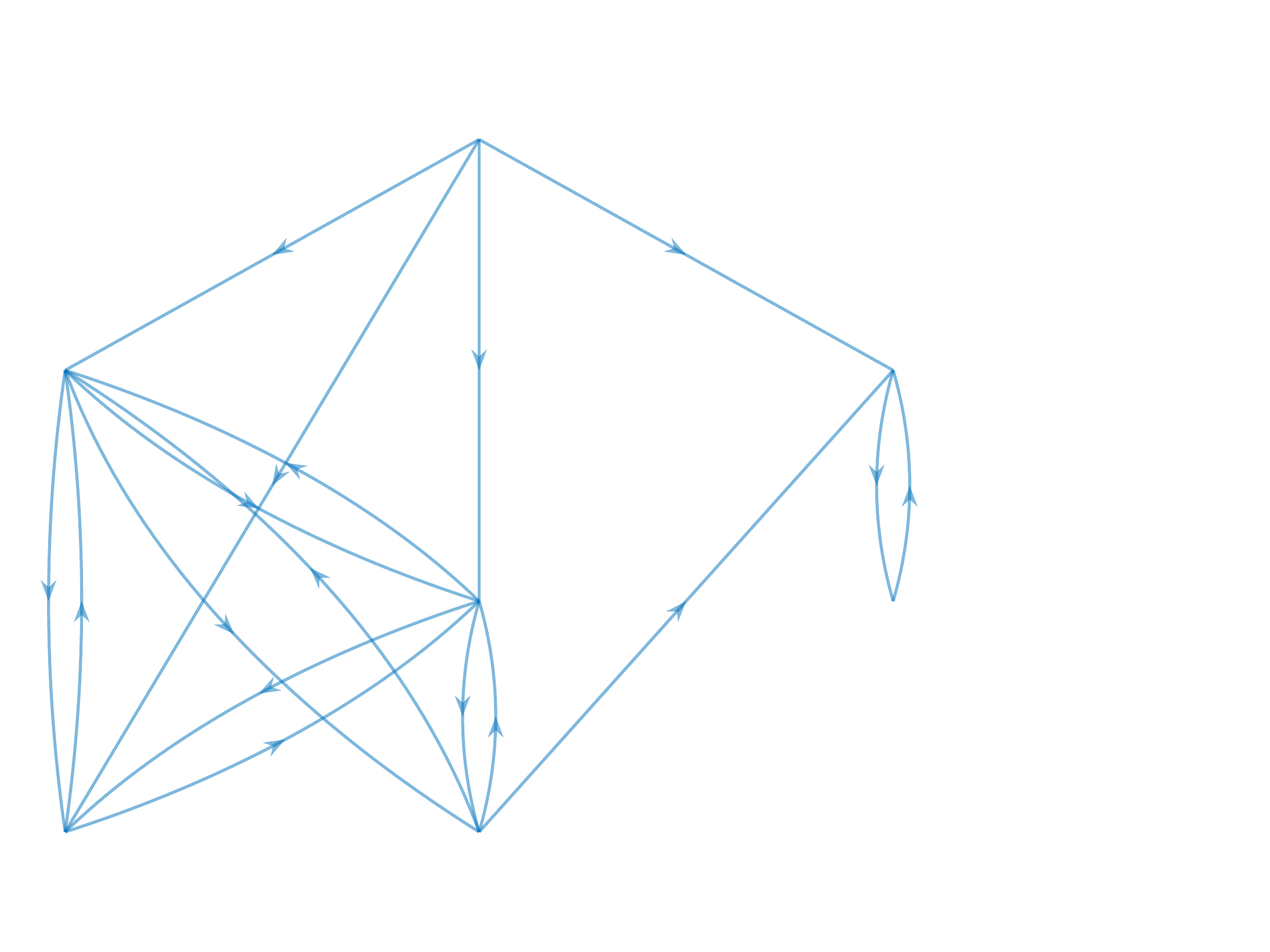
\caption{k=2}
      \label{fig:032}
  \end{subfigure}
  \begin{subfigure}[b]{0.45\linewidth}
 \def\svgwidth{1\linewidth}
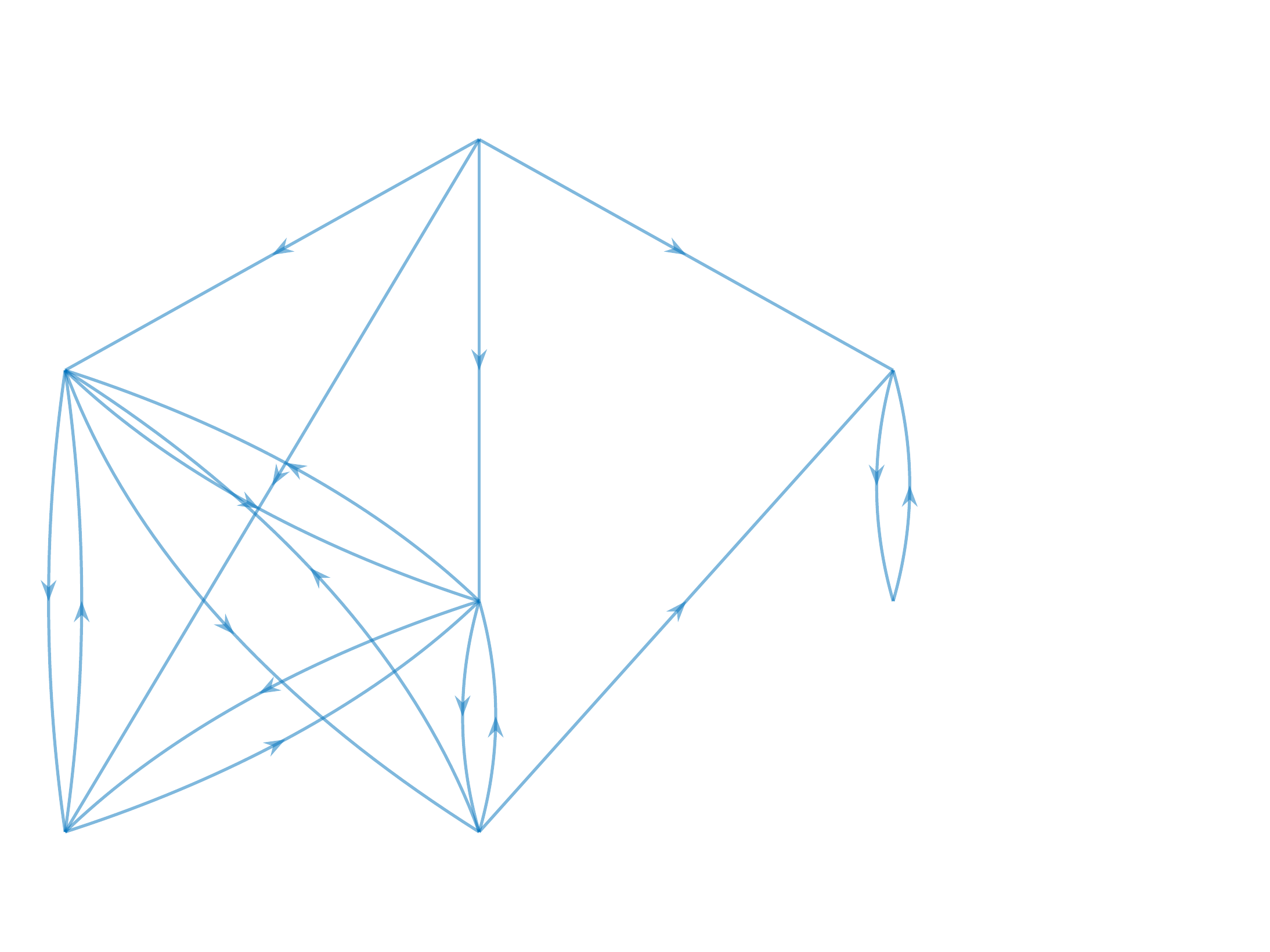
      \caption{k=3}
      \label{fig:033}
  \end{subfigure}
  ~ 
     \begin{subfigure}[b]{0.45\linewidth}
 \def\svgwidth{1\linewidth}
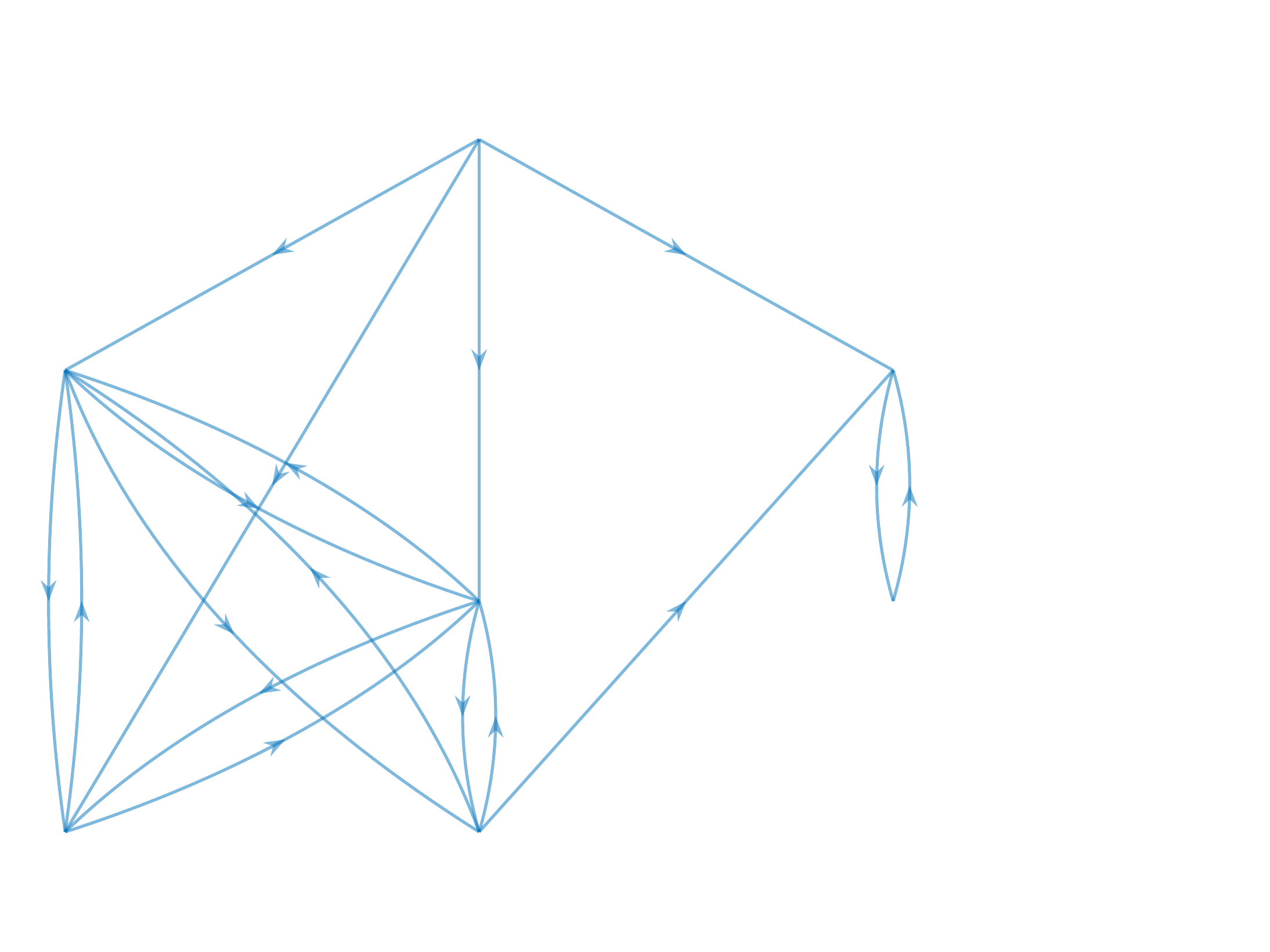
\caption{k=K}
      \label{fig:034}
  \end{subfigure}
      \caption{Target strategy of the \emph{average} graph of Fig. \ref{fig:KRA}. }     
\label{fig:AverageRA}
\end{figure}

We want to compare targeted nodes for different assumptions in speed of the network changes compared to the rumor spreading. Leaving all spread and model parameters the same, we now consider the scenarios of 1) static graph, 2) temporal-switching graph with $K \rightarrow \infty$ and 3) an average graph. The results are displayed in respectively Fig. \ref{fig:03}, \ref{fig:KRA}, and \ref{fig:AverageRA}. It can be seen that the awareness or anti-rumor campaign strategy differs quite a bit depending on the assumptions made on the speed of the network changes and stresses the importance of taking into account the temporal character. All temporal strategies start with targeting node $i=7$ compared to node 1 for the static approach due to knowledge of the changes in influence. Strategies after $k=1$ differ due to different assumptions in speed and knowledge of the network changes and duration.

\subsection{US Air Network}
\begin{figure}
\centering
 \def\svgwidth{0.77\linewidth}
\begingroup%
  \makeatletter%
  \providecommand\color[2][]{%
    \errmessage{(Inkscape) Color is used for the text in Inkscape, but the package 'color.sty' is not loaded}%
    \renewcommand\color[2][]{}%
  }%
  \providecommand\transparent[1]{%
    \errmessage{(Inkscape) Transparency is used (non-zero) for the text in Inkscape, but the package 'transparent.sty' is not loaded}%
    \renewcommand\transparent[1]{}%
  }%
  \providecommand\rotatebox[2]{#2}%
  \newcommand*\fsize{\dimexpr\f@size pt\relax}%
  \newcommand*\lineheight[1]{\fontsize{\fsize}{#1\fsize}\selectfont}%
  \ifx\svgwidth\undefined%
    \setlength{\unitlength}{630bp}%
    \ifx\svgscale\undefined%
      \relax%
    \else%
      \setlength{\unitlength}{\unitlength * \real{\svgscale}}%
    \fi%
  \else%
    \setlength{\unitlength}{\svgwidth}%
  \fi%
  \global\let\svgwidth\undefined%
  \global\let\svgscale\undefined%
  \makeatother%
  \begin{picture}(1,0.75)%
    \lineheight{1}%
    \setlength\tabcolsep{0pt}%
    \put(0,0){\includegraphics[width=\unitlength,page=1]{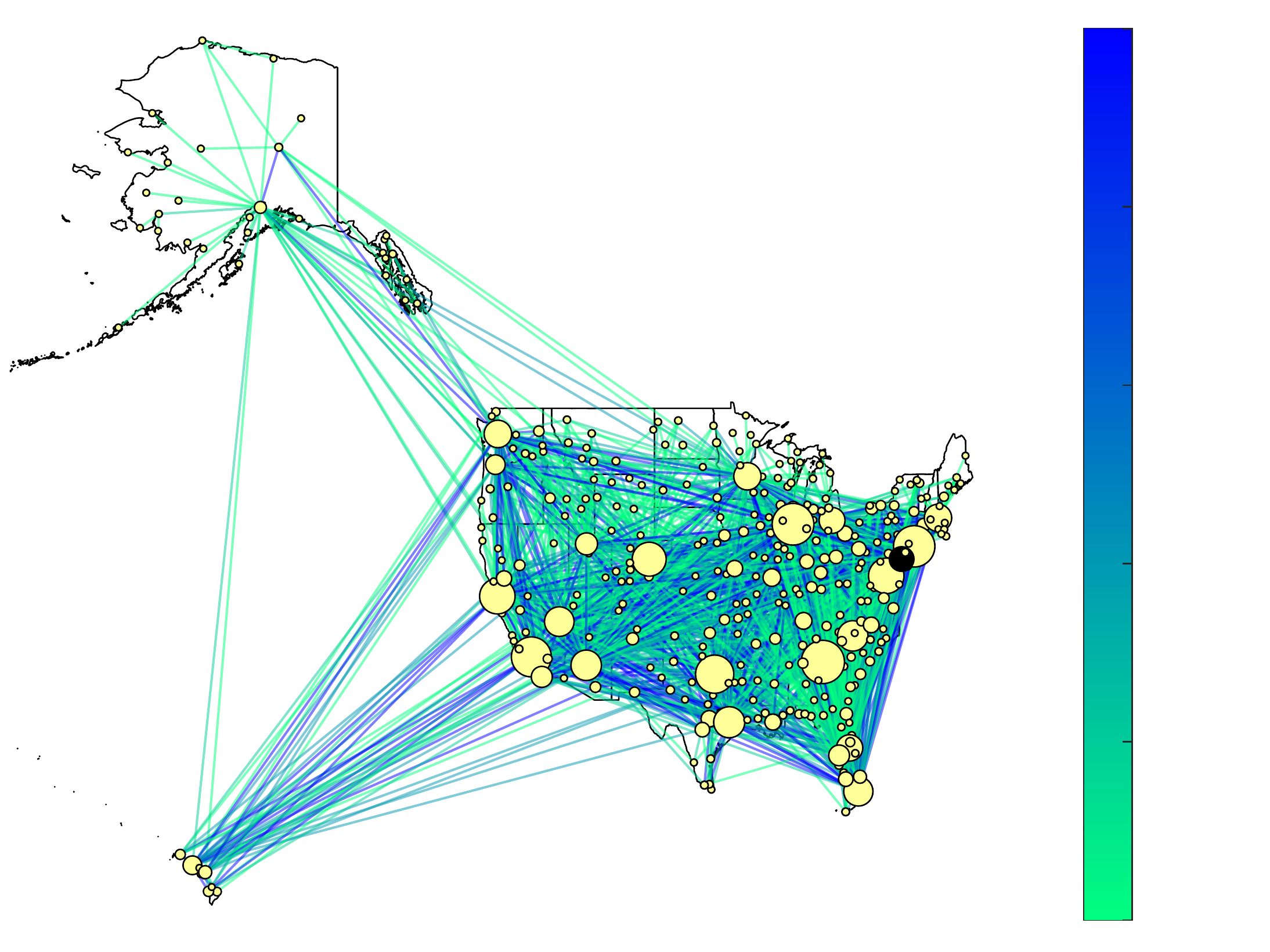}}%
    \put(0.90309524,0.01190476){\makebox(0,0)[lt]{\lineheight{1.25}\smash{\begin{tabular}[t]{l}2\end{tabular}}}}%
    \put(0.90309524,0.15238095){\makebox(0,0)[lt]{\lineheight{1.25}\smash{\begin{tabular}[t]{l}3\end{tabular}}}}%
    \put(0.90309524,0.29285714){\makebox(0,0)[lt]{\lineheight{1.25}\smash{\begin{tabular}[t]{l}4\end{tabular}}}}%
    \put(0.90309524,0.43333333){\makebox(0,0)[lt]{\lineheight{1.25}\smash{\begin{tabular}[t]{l}5\end{tabular}}}}%
    \put(0.90309524,0.57380952){\makebox(0,0)[lt]{\lineheight{1.25}\smash{\begin{tabular}[t]{l}6\end{tabular}}}}%
    \put(0.90309524,0.71428571){\makebox(0,0)[lt]{\lineheight{1.25}\smash{\begin{tabular}[t]{l}7\end{tabular}}}}%
    \put(0.99380952,0.14321464){\rotatebox{90}{\makebox(0,0)[lt]{\lineheight{1.25}\smash{\begin{tabular}[t]{l}Flight days per week\end{tabular}}}}}%
    \put(0,0){\includegraphics[width=\unitlength,page=2]{USPeriodic.pdf}}%
  \end{picture}%
\endgroup%
 
      \caption{Periodic US air transportation network with 359 nodes and 514 - 2097 edges based on \cite{USdata}. Edge color indicates the number of flight days per week with dark blue indicating routes that are flown at least daily. Node marker size indicates cost $c_i$ and node color indicates outbreak probability $\hat{x}_i^1$.}     
\label{fig:USGraph}
\end{figure}

Next, we consider a periodic domestic US air transportation network graph based on the amount of passengers transported in 2014 from \cite{USdata} modified to a periodic flight schedule consisting of $n=359$ nodes and 514 - 2097 edges based on the flight day as visualized in Fig. \ref{fig:USGraph}. All domestic air traffic within the US excluding Puerto Rico, Virgin Islands and Guam is considered, where multiple airports that serve the same city are combined, and routes that have less than 10,000 passengers carried per year are omitted. Number of flight days is determined based on passengers transported and routes are divided into respectively 2, 3, 4 or 7 flight days a week to create a periodic schedule and network graph. 

\begin{figure*}
\centering
\begin{subfigure}[b]{0.3\linewidth}
 \def\svgwidth{1\linewidth}
\begingroup%
  \makeatletter%
  \providecommand\color[2][]{%
    \errmessage{(Inkscape) Color is used for the text in Inkscape, but the package 'color.sty' is not loaded}%
    \renewcommand\color[2][]{}%
  }%
  \providecommand\transparent[1]{%
    \errmessage{(Inkscape) Transparency is used (non-zero) for the text in Inkscape, but the package 'transparent.sty' is not loaded}%
    \renewcommand\transparent[1]{}%
  }%
  \providecommand\rotatebox[2]{#2}%
  \newcommand*\fsize{\dimexpr\f@size pt\relax}%
  \newcommand*\lineheight[1]{\fontsize{\fsize}{#1\fsize}\selectfont}%
  \ifx\svgwidth\undefined%
    \setlength{\unitlength}{630bp}%
    \ifx\svgscale\undefined%
      \relax%
    \else%
      \setlength{\unitlength}{\unitlength * \real{\svgscale}}%
    \fi%
  \else%
    \setlength{\unitlength}{\svgwidth}%
  \fi%
  \global\let\svgwidth\undefined%
  \global\let\svgscale\undefined%
  \makeatother%
  \begin{picture}(1,0.75)%
    \lineheight{1}%
    \setlength\tabcolsep{0pt}%
    \put(0,0){\includegraphics[width=\unitlength,page=1]{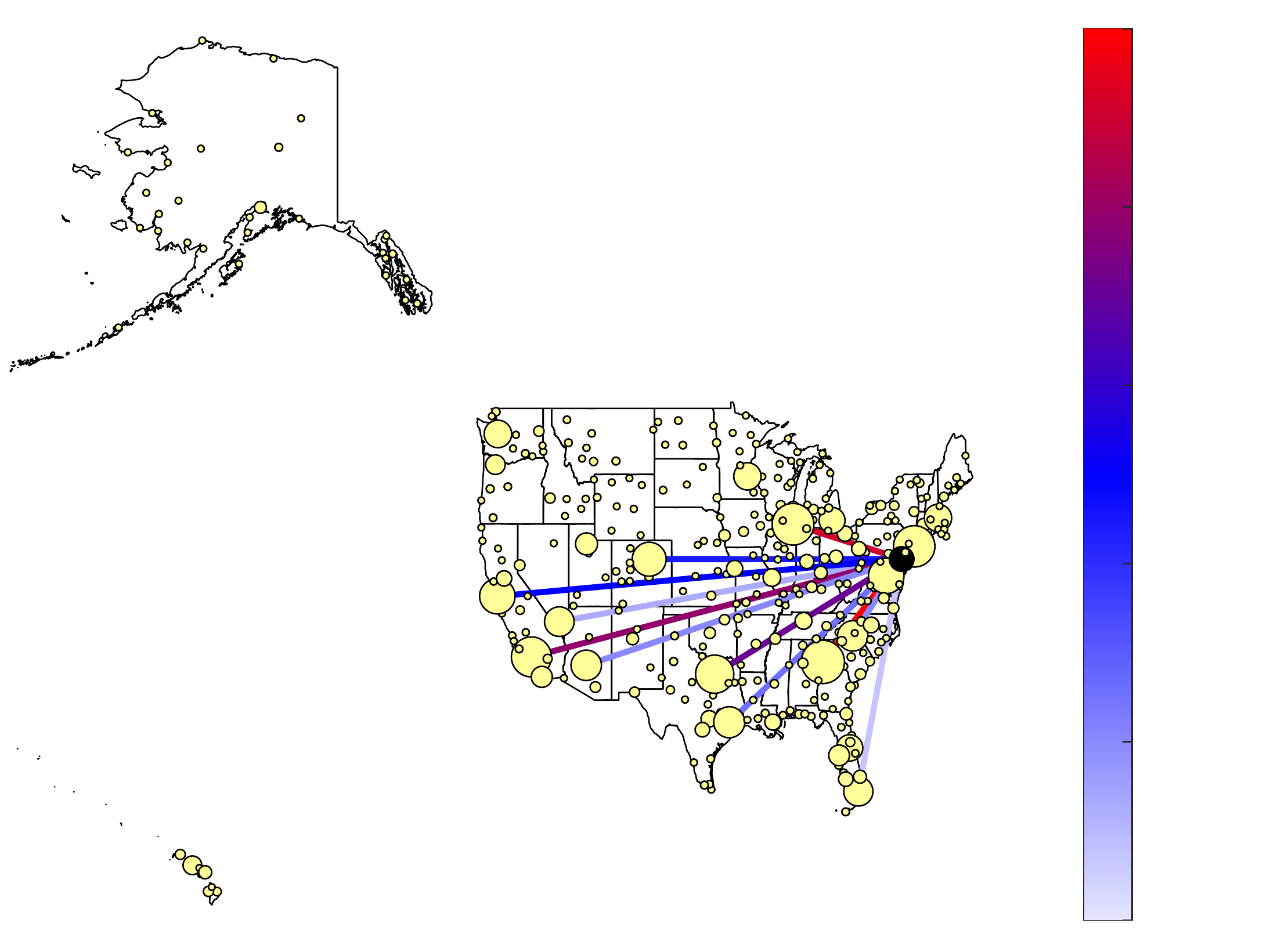}}%
    \put(0.90309524,0.01190476){\makebox(0,0)[lt]{\lineheight{1.25}\smash{\begin{tabular}[t]{l}\scriptsize 0\end{tabular}}}}%
    \put(0.90309524,0.15238095){\makebox(0,0)[lt]{\lineheight{1.25}\smash{\begin{tabular}[t]{l}\scriptsize 0.2\end{tabular}}}}%
    \put(0.90309524,0.29285714){\makebox(0,0)[lt]{\lineheight{1.25}\smash{\begin{tabular}[t]{l}\scriptsize 0.4\end{tabular}}}}%
    \put(0.90309524,0.43333333){\makebox(0,0)[lt]{\lineheight{1.25}\smash{\begin{tabular}[t]{l}\scriptsize 0.6\end{tabular}}}}%
    \put(0.90309524,0.57380952){\makebox(0,0)[lt]{\lineheight{1.25}\smash{\begin{tabular}[t]{l}\scriptsize 0.8\end{tabular}}}}%
    \put(0.90309524,0.71428571){\makebox(0,0)[lt]{\lineheight{1.25}\smash{\begin{tabular}[t]{l}\scriptsize 1\end{tabular}}}}%
    \put(0.99999,0.11738131){\rotatebox{90}{\makebox(0,0)[lt]{\lineheight{1.25}\smash{\begin{tabular}[t]{l}\footnotesize Resource Allocation\end{tabular}}}}}%
    \put(0,0){\includegraphics[width=\unitlength,page=2]{Sunday.pdf}}%
  \end{picture}%
\endgroup%
 
      \caption{Sunday}
      \label{fig:031}
  \end{subfigure}
     \begin{subfigure}[b]{0.3\linewidth}
 \def\svgwidth{1\linewidth}
\begingroup%
  \makeatletter%
  \providecommand\color[2][]{%
    \errmessage{(Inkscape) Color is used for the text in Inkscape, but the package 'color.sty' is not loaded}%
    \renewcommand\color[2][]{}%
  }%
  \providecommand\transparent[1]{%
    \errmessage{(Inkscape) Transparency is used (non-zero) for the text in Inkscape, but the package 'transparent.sty' is not loaded}%
    \renewcommand\transparent[1]{}%
  }%
  \providecommand\rotatebox[2]{#2}%
  \newcommand*\fsize{\dimexpr\f@size pt\relax}%
  \newcommand*\lineheight[1]{\fontsize{\fsize}{#1\fsize}\selectfont}%
  \ifx\svgwidth\undefined%
    \setlength{\unitlength}{630bp}%
    \ifx\svgscale\undefined%
      \relax%
    \else%
      \setlength{\unitlength}{\unitlength * \real{\svgscale}}%
    \fi%
  \else%
    \setlength{\unitlength}{\svgwidth}%
  \fi%
  \global\let\svgwidth\undefined%
  \global\let\svgscale\undefined%
  \makeatother%
  \begin{picture}(1,0.75)%
    \lineheight{1}%
    \setlength\tabcolsep{0pt}%
    \put(0,0){\includegraphics[width=\unitlength,page=1]{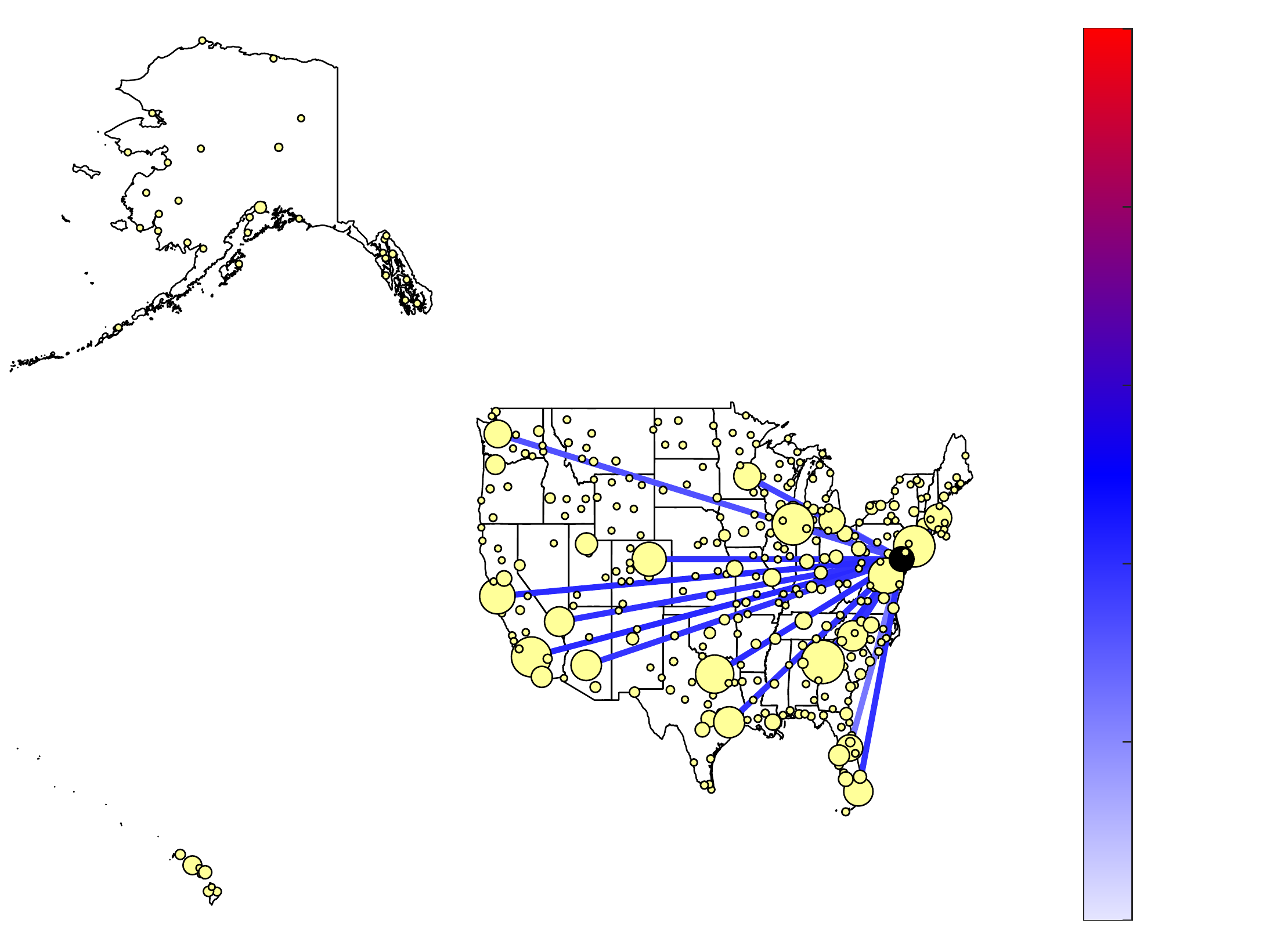}}%
    \put(0.90309524,0.01190476){\makebox(0,0)[lt]{\lineheight{1.25}\smash{\begin{tabular}[t]{l}\scriptsize 0\end{tabular}}}}%
    \put(0.90309524,0.15238095){\makebox(0,0)[lt]{\lineheight{1.25}\smash{\begin{tabular}[t]{l}\scriptsize 0.2\end{tabular}}}}%
    \put(0.90309524,0.29285714){\makebox(0,0)[lt]{\lineheight{1.25}\smash{\begin{tabular}[t]{l}\scriptsize 0.4\end{tabular}}}}%
    \put(0.90309524,0.43333333){\makebox(0,0)[lt]{\lineheight{1.25}\smash{\begin{tabular}[t]{l}\scriptsize 0.6\end{tabular}}}}%
    \put(0.90309524,0.57380952){\makebox(0,0)[lt]{\lineheight{1.25}\smash{\begin{tabular}[t]{l}\scriptsize 0.8\end{tabular}}}}%
    \put(0.90309524,0.71428571){\makebox(0,0)[lt]{\lineheight{1.25}\smash{\begin{tabular}[t]{l}\scriptsize 1\end{tabular}}}}%
    \put(0.99999,0.11738131){\rotatebox{90}{\makebox(0,0)[lt]{\lineheight{1.25}\smash{\begin{tabular}[t]{l}\footnotesize Resource Allocation\end{tabular}}}}}%
    \put(0,0){\includegraphics[width=\unitlength,page=2]{Monday.pdf}}%
  \end{picture}%
\endgroup%
 
\caption{Monday}
      \label{fig:032}
  \end{subfigure}
    \begin{subfigure}[b]{0.3\linewidth}
 \def\svgwidth{1\linewidth}
\begingroup%
  \makeatletter%
  \providecommand\color[2][]{%
    \errmessage{(Inkscape) Color is used for the text in Inkscape, but the package 'color.sty' is not loaded}%
    \renewcommand\color[2][]{}%
  }%
  \providecommand\transparent[1]{%
    \errmessage{(Inkscape) Transparency is used (non-zero) for the text in Inkscape, but the package 'transparent.sty' is not loaded}%
    \renewcommand\transparent[1]{}%
  }%
  \providecommand\rotatebox[2]{#2}%
  \newcommand*\fsize{\dimexpr\f@size pt\relax}%
  \newcommand*\lineheight[1]{\fontsize{\fsize}{#1\fsize}\selectfont}%
  \ifx\svgwidth\undefined%
    \setlength{\unitlength}{630bp}%
    \ifx\svgscale\undefined%
      \relax%
    \else%
      \setlength{\unitlength}{\unitlength * \real{\svgscale}}%
    \fi%
  \else%
    \setlength{\unitlength}{\svgwidth}%
  \fi%
  \global\let\svgwidth\undefined%
  \global\let\svgscale\undefined%
  \makeatother%
  \begin{picture}(1,0.75)%
    \lineheight{1}%
    \setlength\tabcolsep{0pt}%
    \put(0,0){\includegraphics[width=\unitlength,page=1]{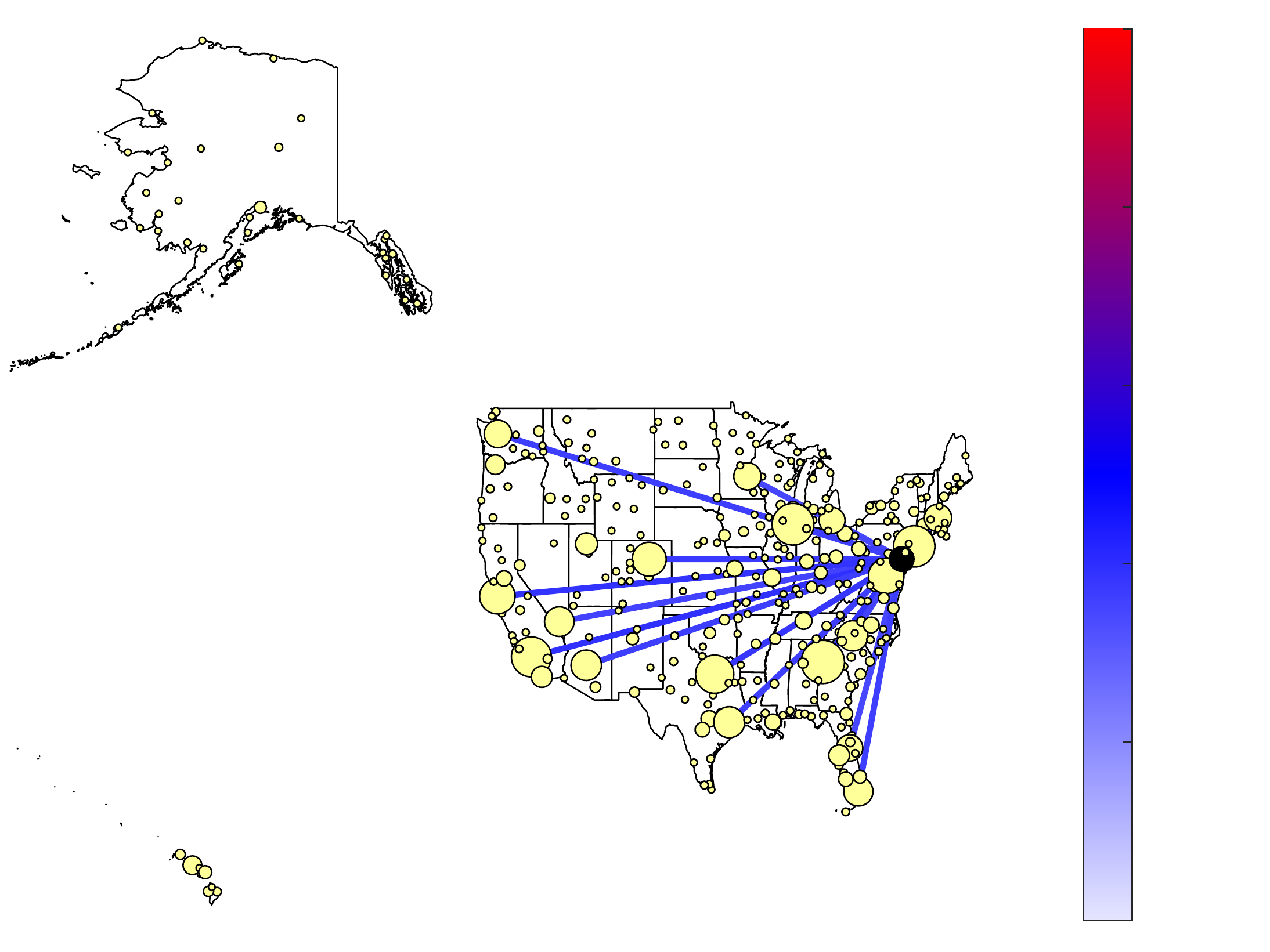}}%
    \put(0.90309524,0.01190476){\makebox(0,0)[lt]{\lineheight{1.25}\smash{\begin{tabular}[t]{l}\scriptsize 0\end{tabular}}}}%
    \put(0.90309524,0.15238095){\makebox(0,0)[lt]{\lineheight{1.25}\smash{\begin{tabular}[t]{l}\scriptsize 0.2\end{tabular}}}}%
    \put(0.90309524,0.29285714){\makebox(0,0)[lt]{\lineheight{1.25}\smash{\begin{tabular}[t]{l}\scriptsize 0.4\end{tabular}}}}%
    \put(0.90309524,0.43333333){\makebox(0,0)[lt]{\lineheight{1.25}\smash{\begin{tabular}[t]{l}\scriptsize 0.6\end{tabular}}}}%
    \put(0.90309524,0.57380952){\makebox(0,0)[lt]{\lineheight{1.25}\smash{\begin{tabular}[t]{l}\scriptsize 0.8\end{tabular}}}}%
    \put(0.90309524,0.71428571){\makebox(0,0)[lt]{\lineheight{1.25}\smash{\begin{tabular}[t]{l}\scriptsize 1\end{tabular}}}}%
    \put(0.99999,0.11738131){\rotatebox{90}{\makebox(0,0)[lt]{\lineheight{1.25}\smash{\begin{tabular}[t]{l}\footnotesize Resource Allocation\end{tabular}}}}}%
    \put(0,0){\includegraphics[width=\unitlength,page=2]{Tuesday.pdf}}%
  \end{picture}%
\endgroup%
 
\caption{Tuesday}
      \label{fig:032}
  \end{subfigure}
  \begin{subfigure}[b]{0.3\linewidth}
 \def\svgwidth{1\linewidth}
\begingroup%
  \makeatletter%
  \providecommand\color[2][]{%
    \errmessage{(Inkscape) Color is used for the text in Inkscape, but the package 'color.sty' is not loaded}%
    \renewcommand\color[2][]{}%
  }%
  \providecommand\transparent[1]{%
    \errmessage{(Inkscape) Transparency is used (non-zero) for the text in Inkscape, but the package 'transparent.sty' is not loaded}%
    \renewcommand\transparent[1]{}%
  }%
  \providecommand\rotatebox[2]{#2}%
  \newcommand*\fsize{\dimexpr\f@size pt\relax}%
  \newcommand*\lineheight[1]{\fontsize{\fsize}{#1\fsize}\selectfont}%
  \ifx\svgwidth\undefined%
    \setlength{\unitlength}{630bp}%
    \ifx\svgscale\undefined%
      \relax%
    \else%
      \setlength{\unitlength}{\unitlength * \real{\svgscale}}%
    \fi%
  \else%
    \setlength{\unitlength}{\svgwidth}%
  \fi%
  \global\let\svgwidth\undefined%
  \global\let\svgscale\undefined%
  \makeatother%
  \begin{picture}(1,0.75)%
    \lineheight{1}%
    \setlength\tabcolsep{0pt}%
    \put(0,0){\includegraphics[width=\unitlength,page=1]{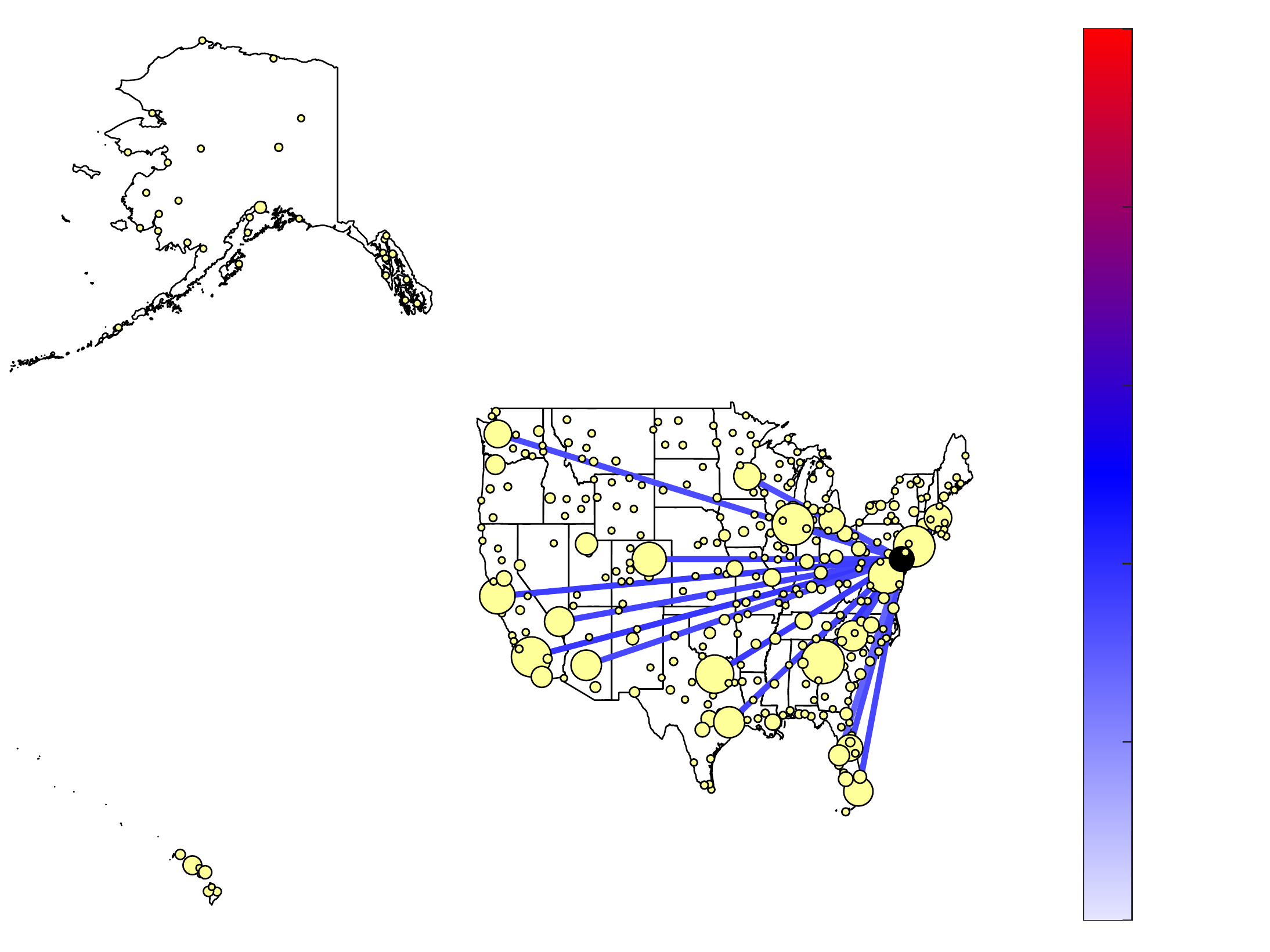}}%
    \put(0.90309524,0.01190476){\makebox(0,0)[lt]{\lineheight{1.25}\smash{\begin{tabular}[t]{l}\scriptsize 0\end{tabular}}}}%
    \put(0.90309524,0.15238095){\makebox(0,0)[lt]{\lineheight{1.25}\smash{\begin{tabular}[t]{l}\scriptsize 0.2\end{tabular}}}}%
    \put(0.90309524,0.29285714){\makebox(0,0)[lt]{\lineheight{1.25}\smash{\begin{tabular}[t]{l}\scriptsize 0.4\end{tabular}}}}%
    \put(0.90309524,0.43333333){\makebox(0,0)[lt]{\lineheight{1.25}\smash{\begin{tabular}[t]{l}\scriptsize 0.6\end{tabular}}}}%
    \put(0.90309524,0.57380952){\makebox(0,0)[lt]{\lineheight{1.25}\smash{\begin{tabular}[t]{l}\scriptsize 0.8\end{tabular}}}}%
    \put(0.90309524,0.71428571){\makebox(0,0)[lt]{\lineheight{1.25}\smash{\begin{tabular}[t]{l}\scriptsize 1\end{tabular}}}}%
    \put(0.99999,0.11738131){\rotatebox{90}{\makebox(0,0)[lt]{\lineheight{1.25}\smash{\begin{tabular}[t]{l}\footnotesize Resource Allocation\end{tabular}}}}}%
    \put(0,0){\includegraphics[width=\unitlength,page=2]{Wednesday.pdf}}%
  \end{picture}%
\endgroup%
 
      \caption{Wednesday}
      \label{fig:033}
  \end{subfigure}
     \begin{subfigure}[b]{0.3\linewidth}
 \def\svgwidth{1\linewidth}
\begingroup%
  \makeatletter%
  \providecommand\color[2][]{%
    \errmessage{(Inkscape) Color is used for the text in Inkscape, but the package 'color.sty' is not loaded}%
    \renewcommand\color[2][]{}%
  }%
  \providecommand\transparent[1]{%
    \errmessage{(Inkscape) Transparency is used (non-zero) for the text in Inkscape, but the package 'transparent.sty' is not loaded}%
    \renewcommand\transparent[1]{}%
  }%
  \providecommand\rotatebox[2]{#2}%
  \newcommand*\fsize{\dimexpr\f@size pt\relax}%
  \newcommand*\lineheight[1]{\fontsize{\fsize}{#1\fsize}\selectfont}%
  \ifx\svgwidth\undefined%
    \setlength{\unitlength}{630bp}%
    \ifx\svgscale\undefined%
      \relax%
    \else%
      \setlength{\unitlength}{\unitlength * \real{\svgscale}}%
    \fi%
  \else%
    \setlength{\unitlength}{\svgwidth}%
  \fi%
  \global\let\svgwidth\undefined%
  \global\let\svgscale\undefined%
  \makeatother%
  \begin{picture}(1,0.75)%
    \lineheight{1}%
    \setlength\tabcolsep{0pt}%
    \put(0,0){\includegraphics[width=\unitlength,page=1]{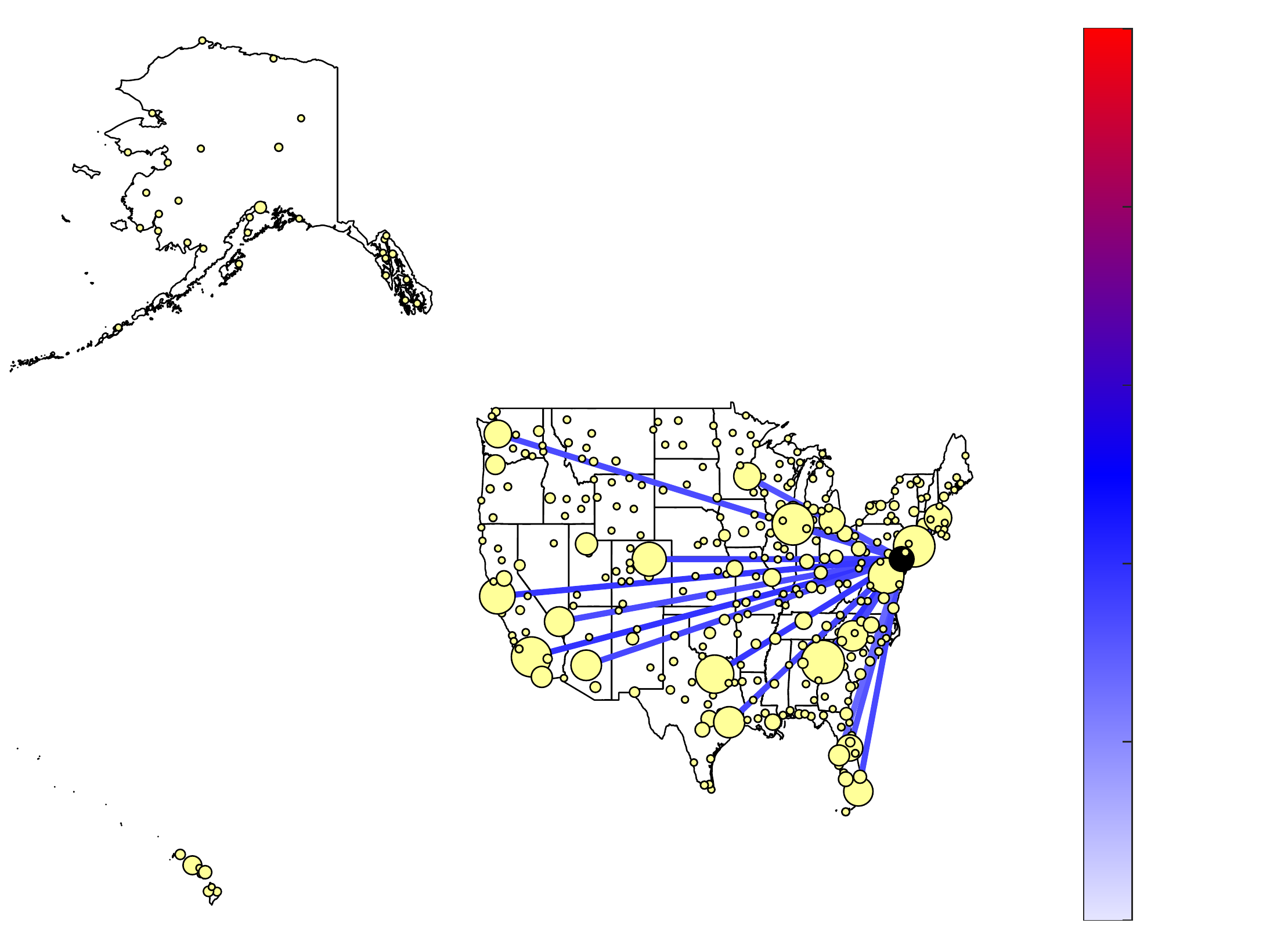}}%
    \put(0.90309524,0.01190476){\makebox(0,0)[lt]{\lineheight{1.25}\smash{\begin{tabular}[t]{l}\scriptsize 0\end{tabular}}}}%
    \put(0.90309524,0.15238095){\makebox(0,0)[lt]{\lineheight{1.25}\smash{\begin{tabular}[t]{l}\scriptsize 0.2\end{tabular}}}}%
    \put(0.90309524,0.29285714){\makebox(0,0)[lt]{\lineheight{1.25}\smash{\begin{tabular}[t]{l}\scriptsize 0.4\end{tabular}}}}%
    \put(0.90309524,0.43333333){\makebox(0,0)[lt]{\lineheight{1.25}\smash{\begin{tabular}[t]{l}\scriptsize 0.6\end{tabular}}}}%
    \put(0.90309524,0.57380952){\makebox(0,0)[lt]{\lineheight{1.25}\smash{\begin{tabular}[t]{l}\scriptsize 0.8\end{tabular}}}}%
    \put(0.90309524,0.71428571){\makebox(0,0)[lt]{\lineheight{1.25}\smash{\begin{tabular}[t]{l}\scriptsize 1\end{tabular}}}}%
    \put(0.99999,0.11738131){\rotatebox{90}{\makebox(0,0)[lt]{\lineheight{1.25}\smash{\begin{tabular}[t]{l}\footnotesize Resource Allocation\end{tabular}}}}}%
    \put(0,0){\includegraphics[width=\unitlength,page=2]{Thursday.pdf}}%
  \end{picture}%
\endgroup%
 
\caption{Thursday}
      \label{fig:034}
  \end{subfigure}
    \begin{subfigure}[b]{0.3\linewidth}
 \def\svgwidth{1\linewidth}
\begingroup%
  \makeatletter%
  \providecommand\color[2][]{%
    \errmessage{(Inkscape) Color is used for the text in Inkscape, but the package 'color.sty' is not loaded}%
    \renewcommand\color[2][]{}%
  }%
  \providecommand\transparent[1]{%
    \errmessage{(Inkscape) Transparency is used (non-zero) for the text in Inkscape, but the package 'transparent.sty' is not loaded}%
    \renewcommand\transparent[1]{}%
  }%
  \providecommand\rotatebox[2]{#2}%
  \newcommand*\fsize{\dimexpr\f@size pt\relax}%
  \newcommand*\lineheight[1]{\fontsize{\fsize}{#1\fsize}\selectfont}%
  \ifx\svgwidth\undefined%
    \setlength{\unitlength}{630bp}%
    \ifx\svgscale\undefined%
      \relax%
    \else%
      \setlength{\unitlength}{\unitlength * \real{\svgscale}}%
    \fi%
  \else%
    \setlength{\unitlength}{\svgwidth}%
  \fi%
  \global\let\svgwidth\undefined%
  \global\let\svgscale\undefined%
  \makeatother%
  \begin{picture}(1,0.75)%
    \lineheight{1}%
    \setlength\tabcolsep{0pt}%
    \put(0,0){\includegraphics[width=\unitlength,page=1]{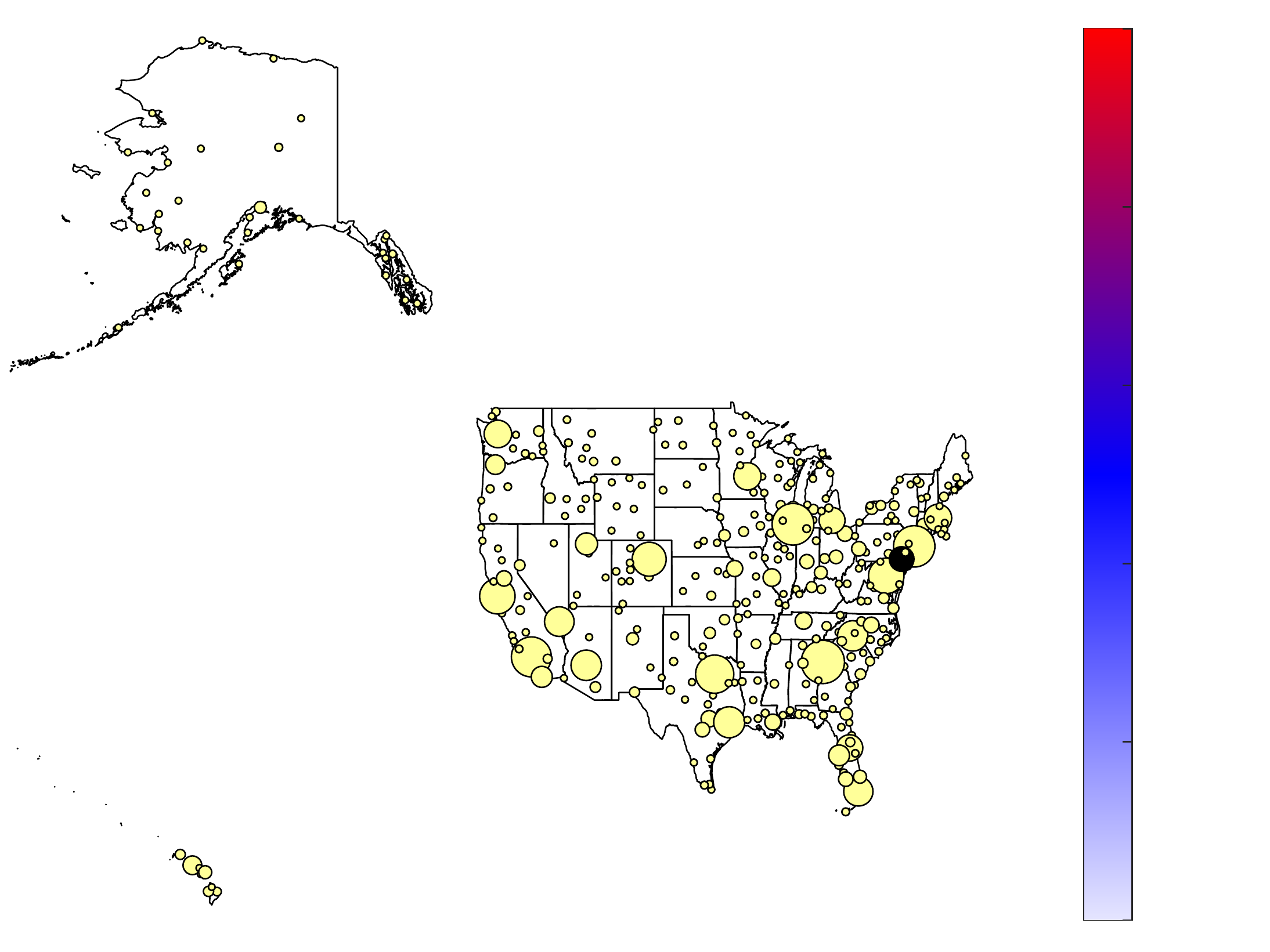}}%
    \put(0.90309524,0.01190476){\makebox(0,0)[lt]{\lineheight{1.25}\smash{\begin{tabular}[t]{l}\scriptsize 0\end{tabular}}}}%
    \put(0.90309524,0.15238095){\makebox(0,0)[lt]{\lineheight{1.25}\smash{\begin{tabular}[t]{l}\scriptsize 0.2\end{tabular}}}}%
    \put(0.90309524,0.29285714){\makebox(0,0)[lt]{\lineheight{1.25}\smash{\begin{tabular}[t]{l}\scriptsize 0.4\end{tabular}}}}%
    \put(0.90309524,0.43333333){\makebox(0,0)[lt]{\lineheight{1.25}\smash{\begin{tabular}[t]{l}\scriptsize 0.6\end{tabular}}}}%
    \put(0.90309524,0.57380952){\makebox(0,0)[lt]{\lineheight{1.25}\smash{\begin{tabular}[t]{l}\scriptsize 0.8\end{tabular}}}}%
    \put(0.90309524,0.71428571){\makebox(0,0)[lt]{\lineheight{1.25}\smash{\begin{tabular}[t]{l}\scriptsize 1\end{tabular}}}}%
    \put(0.99999,0.11738131){\rotatebox{90}{\makebox(0,0)[lt]{\lineheight{1.25}\smash{\begin{tabular}[t]{l}\footnotesize Resource Allocation\end{tabular}}}}}%
    \put(0,0){\includegraphics[width=\unitlength,page=2]{Friday.pdf}}%
  \end{picture}%
\endgroup%
 
\caption{Friday, Saturday}
      \label{fig:032}
  \end{subfigure}
      \caption{Daily reduction of passenger numbers on respectively 13, 18, 18, 19, 19, 0 and 0 routes for $\Gamma^k=5$ and $\Gamma_{\text{tot}}=25$ to repress an outbreak in Philadelphia.}     
\label{fig:USRA}
\end{figure*}

We now model an outbreak of SARS in Philadelphia. Spreading parameters are based on \cite{chowell2004model} and we take a spreading rate of $\beta=0.25$ per day and a recovery rate of $\delta=0.0631$. The spreading rate $\beta_{ij}$ can now be controlled by reducing passenger numbers and flights on particular routes. Furthermore, we take a cost $c_i$ based on the normalized number of passengers served per airport to represent the impact of the disease reaching a given city node. We now set a daily budget of $\Gamma^k=5$, with an overall budget of $\Gamma_{\text{tot}}=25$, $h=0.02$ and $\alpha=0.9$, and obtain the daily reduced schedules as visualized in Fig. \ref{fig:USRA}. Passengers numbers are reduced on respectively 13, 18, 18, 19, and 19 routes for the first 5 days, using the maximum budget in the first time steps to maximally oppress the outbreak. 

\subsection{Wildfire Example}
\begin{figure}
\centering
\begin{subfigure}[b]{0.85\linewidth}
\def\svgwidth{1\textwidth}
\includegraphics[width=1\linewidth]{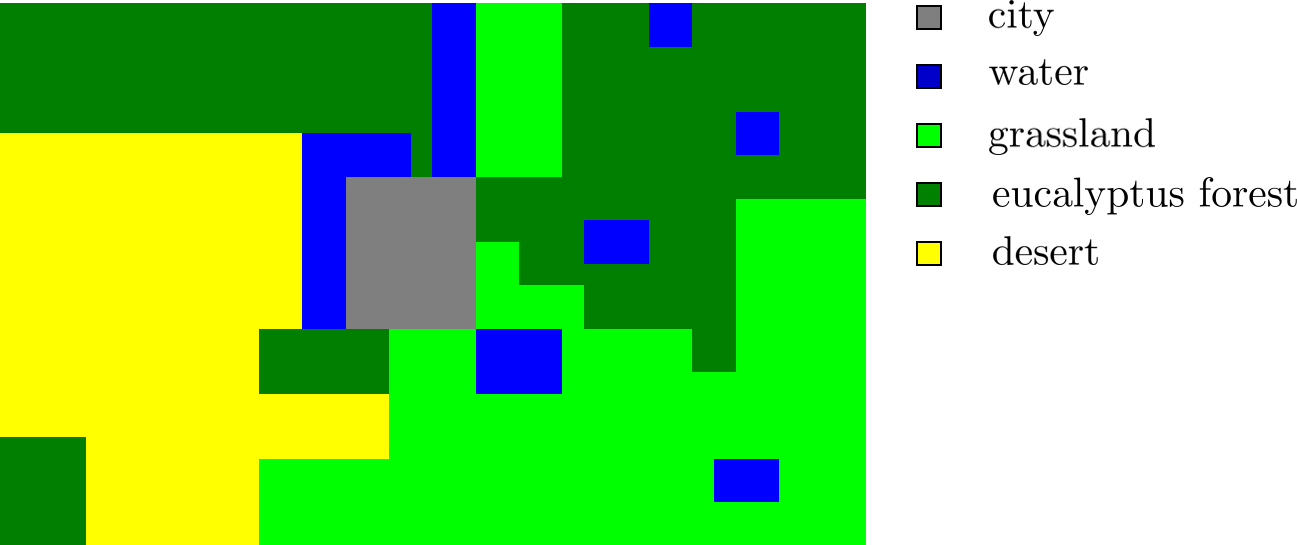}
      \caption{Vegetation Map}
\label{fig:Landscape}
\end{subfigure}
     \begin{subfigure}[b]{0.85\linewidth}
\def\svgwidth{0.79\textwidth}
\begingroup%
  \makeatletter%
  \providecommand\color[2][]{%
    \errmessage{(Inkscape) Color is used for the text in Inkscape, but the package 'color.sty' is not loaded}%
    \renewcommand\color[2][]{}%
  }%
  \providecommand\transparent[1]{%
    \errmessage{(Inkscape) Transparency is used (non-zero) for the text in Inkscape, but the package 'transparent.sty' is not loaded}%
    \renewcommand\transparent[1]{}%
  }%
  \providecommand\rotatebox[2]{#2}%
  \newcommand*\fsize{\dimexpr\f@size pt\relax}%
  \newcommand*\lineheight[1]{\fontsize{\fsize}{#1\fsize}\selectfont}%
  \ifx\svgwidth\undefined%
    \setlength{\unitlength}{516.21487427bp}%
    \ifx\svgscale\undefined%
      \relax%
    \else%
      \setlength{\unitlength}{\unitlength * \real{\svgscale}}%
    \fi%
  \else%
    \setlength{\unitlength}{\svgwidth}%
  \fi%
  \global\let\svgwidth\undefined%
  \global\let\svgscale\undefined%
  \makeatother%
  \begin{picture}(1,0.54687012)%
    \lineheight{1}%
    \setlength\tabcolsep{0pt}%
    \put(0,0){\includegraphics[width=\unitlength,page=1]{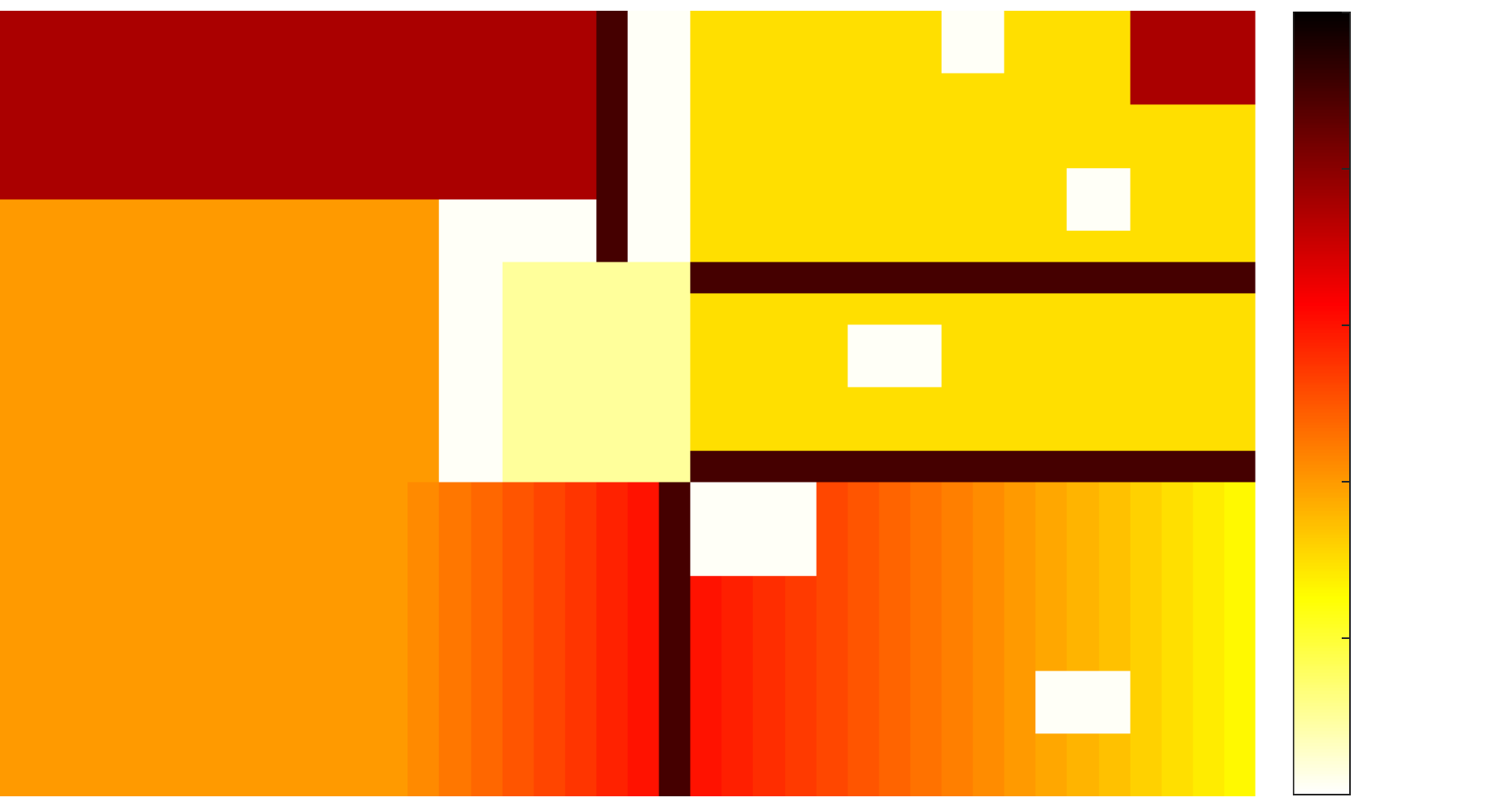}}%
    \put(0.91957826,0.00039585){\makebox(0,0)[lt]{\lineheight{1.25}\smash{\begin{tabular}[t]{l}\scriptsize0\end{tabular}}}}%
    \put(0.91957826,0.10587519){\makebox(0,0)[lt]{\lineheight{1.25}\smash{\begin{tabular}[t]{l}\tiny 0.2\end{tabular}}}}%
    \put(0.91957826,0.21135452){\makebox(0,0)[lt]{\lineheight{1.25}\smash{\begin{tabular}[t]{l}\tiny 0.4\end{tabular}}}}%
    \put(0.91957826,0.31683385){\makebox(0,0)[lt]{\lineheight{1.25}\smash{\begin{tabular}[t]{l}\tiny 0.6\end{tabular}}}}%
    \put(0.91957826,0.42231318){\makebox(0,0)[lt]{\lineheight{1.25}\smash{\begin{tabular}[t]{l}\tiny 0.8\end{tabular}}}}%
    \put(0.91957826,0.52779251){\makebox(0,0)[lt]{\lineheight{1.25}\smash{\begin{tabular}[t]{l}\scriptsize1\end{tabular}}}}%
    \put(0.99999,0.04457196){\rotatebox{90}{\makebox(0,0)[lt]{\lineheight{1.25}\smash{\begin{tabular}[t]{l}\small Outbreak Probability\end{tabular}}}}}%
    \put(0,0){\includegraphics[width=\unitlength,page=2]{Ex2L.pdf}}%
  \end{picture}%
\endgroup%
   
      \caption{Outbreak Probability}
      \label{fig:LM}
  \end{subfigure}
      \caption{Fictional landscape with different area types, represented as a grid with $n=1000$ nodes with its corresponding estimated outbreak probability $\hat{x}_i^1$.}
\label{fig:53}
\end{figure}

Let us consider the fictional landscape given in Fig. \ref{fig:Landscape} consisting of different vegetation types, a city and water. We represent this landscape as a temporal graph $\mathcal{G}^k (\mathcal{V},\mathcal{E}^k)$ with $n=1000$ nodes, where the edge set $\mathcal{E}^k$ connects each node to its neighboring 8 nodes, i.e. fire can spread horizontally, vertically and diagonally, but the spreading values associated with each edge can change each time step in line with changing weather conditions and forecasts. Besides weather forecast, the spreading rates are determined by the vegetation type following \cite{Karafyllidis1997a} and \cite{Alexandridis2008a}, where $\beta=\beta_{b}\beta_{veg}\beta_{w}$, consists of the baseline spreading rate $\beta_{b}=0.5$, the vegetation factor $ \beta_{veg}=0, 0.1, 1$ and $1.4$ for respectively water, desert, grassland and eucalypt forest and the weather effect $\beta_{w}$. Finally, the spreading rate is corrected for the spread between diagonally connected nodes. We set the recovery rate $\delta=0.5$ for all nodes $i \in \mathcal{V}$.

Next, we take the cost $c_i=1$ for the city nodes with $c_i=0.001$ elsewhere and outbreak probability $\hat{x}_{i}^1$ as depicted in Fig. \ref{fig:LM}. This could reflect a higher cost due to risk of human life and property and higher outbreak probability of fires near roads.  

\begin{figure*}
\centering
\begin{subfigure}[b]{0.3\linewidth}
 \def\svgwidth{1\linewidth}
\begingroup%
  \makeatletter%
  \providecommand\color[2][]{%
    \errmessage{(Inkscape) Color is used for the text in Inkscape, but the package 'color.sty' is not loaded}%
    \renewcommand\color[2][]{}%
  }%
  \providecommand\transparent[1]{%
    \errmessage{(Inkscape) Transparency is used (non-zero) for the text in Inkscape, but the package 'transparent.sty' is not loaded}%
    \renewcommand\transparent[1]{}%
  }%
  \providecommand\rotatebox[2]{#2}%
  \newcommand*\fsize{\dimexpr\f@size pt\relax}%
  \newcommand*\lineheight[1]{\fontsize{\fsize}{#1\fsize}\selectfont}%
  \ifx\svgwidth\undefined%
    \setlength{\unitlength}{603.58744812bp}%
    \ifx\svgscale\undefined%
      \relax%
    \else%
      \setlength{\unitlength}{\unitlength * \real{\svgscale}}%
    \fi%
  \else%
    \setlength{\unitlength}{\svgwidth}%
  \fi%
  \global\let\svgwidth\undefined%
  \global\let\svgscale\undefined%
  \makeatother%
  \begin{picture}(1,0.57417196)%
    \lineheight{1}%
    \setlength\tabcolsep{0pt}%
    \put(0,0){\includegraphics[width=\unitlength,page=1]{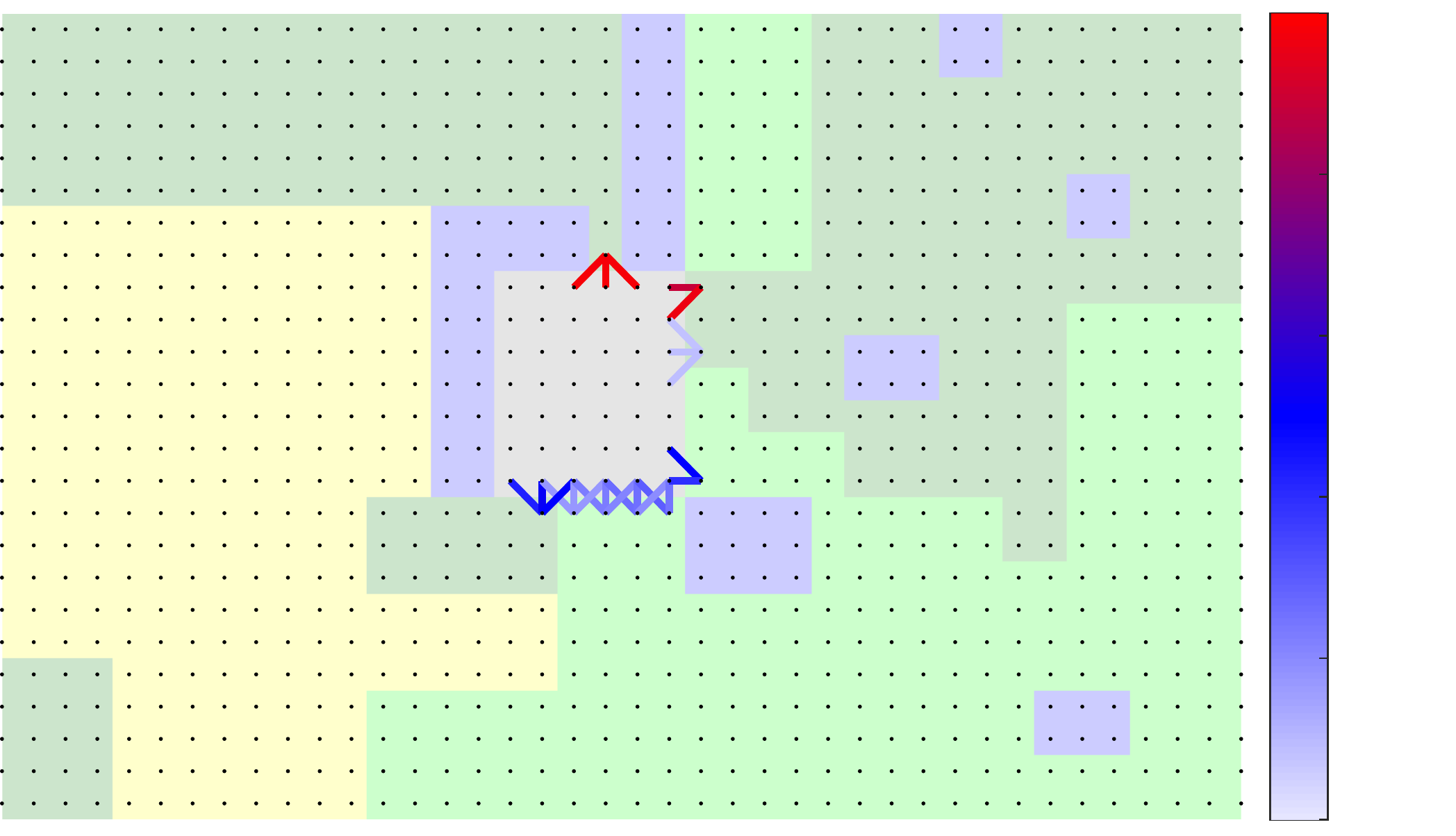}}%
    \put(0.92250924,0.00040626){\makebox(0,0)[lt]{\lineheight{1.25}\smash{\begin{tabular}[t]{l}\tiny 0\end{tabular}}}}%
    \put(0.92250924,0.11124356){\makebox(0,0)[lt]{\lineheight{1.25}\smash{\begin{tabular}[t]{l}\tiny .2\end{tabular}}}}%
    \put(0.92250924,0.22208085){\makebox(0,0)[lt]{\lineheight{1.25}\smash{\begin{tabular}[t]{l}\tiny .4\end{tabular}}}}%
    \put(0.92250924,0.33291815){\makebox(0,0)[lt]{\lineheight{1.25}\smash{\begin{tabular}[t]{l}\tiny .6\end{tabular}}}}%
    \put(0.92250924,0.44375544){\makebox(0,0)[lt]{\lineheight{1.25}\smash{\begin{tabular}[t]{l}\tiny .8\end{tabular}}}}%
    \put(0.92250924,0.55459274){\makebox(0,0)[lt]{\lineheight{1.25}\smash{\begin{tabular}[t]{l}\tiny 1\end{tabular}}}}%
    \put(0.99999,0.0809704){\rotatebox{90}{\makebox(0,0)[lt]{\lineheight{1.25}\smash{\begin{tabular}[t]{l}\scriptsize Resource allocation\end{tabular}}}}}%
    \put(0,0){\includegraphics[width=\unitlength,page=2]{Fire1.pdf}}%
  \end{picture}%
\endgroup%
 
      \caption{k=1}
      \label{fig:031}
  \end{subfigure}
     \begin{subfigure}[b]{0.3\linewidth}
 \def\svgwidth{1\linewidth}
\begingroup%
  \makeatletter%
  \providecommand\color[2][]{%
    \errmessage{(Inkscape) Color is used for the text in Inkscape, but the package 'color.sty' is not loaded}%
    \renewcommand\color[2][]{}%
  }%
  \providecommand\transparent[1]{%
    \errmessage{(Inkscape) Transparency is used (non-zero) for the text in Inkscape, but the package 'transparent.sty' is not loaded}%
    \renewcommand\transparent[1]{}%
  }%
  \providecommand\rotatebox[2]{#2}%
  \newcommand*\fsize{\dimexpr\f@size pt\relax}%
  \newcommand*\lineheight[1]{\fontsize{\fsize}{#1\fsize}\selectfont}%
  \ifx\svgwidth\undefined%
    \setlength{\unitlength}{603.58744812bp}%
    \ifx\svgscale\undefined%
      \relax%
    \else%
      \setlength{\unitlength}{\unitlength * \real{\svgscale}}%
    \fi%
  \else%
    \setlength{\unitlength}{\svgwidth}%
  \fi%
  \global\let\svgwidth\undefined%
  \global\let\svgscale\undefined%
  \makeatother%
  \begin{picture}(1,0.57417196)%
    \lineheight{1}%
    \setlength\tabcolsep{0pt}%
    \put(0,0){\includegraphics[width=\unitlength,page=1]{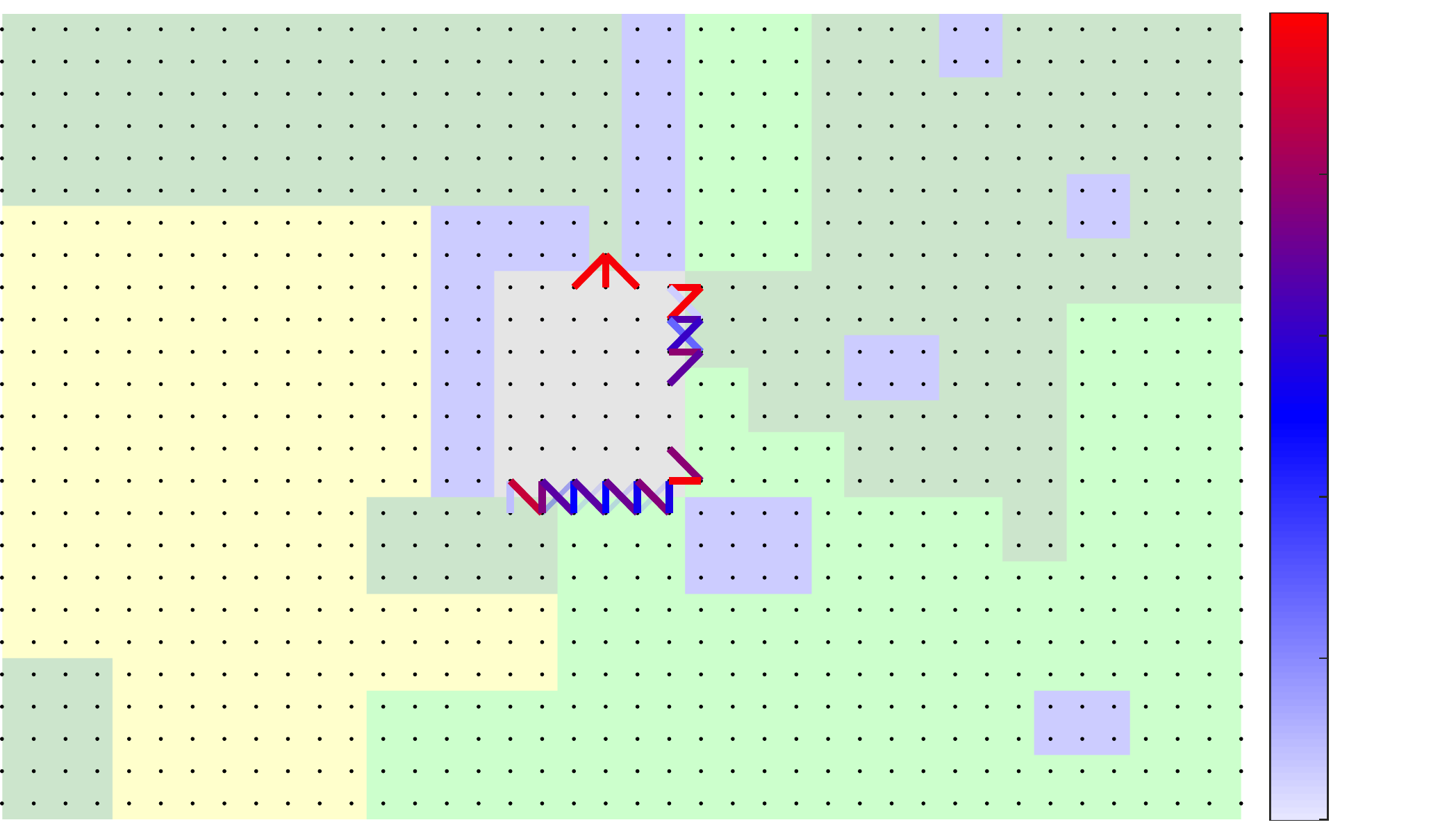}}%
    \put(0.92250924,0.00040626){\makebox(0,0)[lt]{\lineheight{1.25}\smash{\begin{tabular}[t]{l}\tiny 0\end{tabular}}}}%
    \put(0.92250924,0.11124356){\makebox(0,0)[lt]{\lineheight{1.25}\smash{\begin{tabular}[t]{l}\tiny .2\end{tabular}}}}%
    \put(0.92250924,0.22208085){\makebox(0,0)[lt]{\lineheight{1.25}\smash{\begin{tabular}[t]{l}\tiny .4\end{tabular}}}}%
    \put(0.92250924,0.33291815){\makebox(0,0)[lt]{\lineheight{1.25}\smash{\begin{tabular}[t]{l}\tiny .6\end{tabular}}}}%
    \put(0.92250924,0.44375544){\makebox(0,0)[lt]{\lineheight{1.25}\smash{\begin{tabular}[t]{l}\tiny .8\end{tabular}}}}%
    \put(0.92250924,0.55459274){\makebox(0,0)[lt]{\lineheight{1.25}\smash{\begin{tabular}[t]{l}\tiny 1\end{tabular}}}}%
    \put(0.99999,0.0809704){\rotatebox{90}{\makebox(0,0)[lt]{\lineheight{1.25}\smash{\begin{tabular}[t]{l}\scriptsize Resource allocation\end{tabular}}}}}%
    \put(0,0){\includegraphics[width=\unitlength,page=2]{Fire2.pdf}}%
  \end{picture}%
\endgroup%
 
\caption{k=2}
      \label{fig:032}
  \end{subfigure}
    \begin{subfigure}[b]{0.3\linewidth}
 \def\svgwidth{1\linewidth}
\begingroup%
  \makeatletter%
  \providecommand\color[2][]{%
    \errmessage{(Inkscape) Color is used for the text in Inkscape, but the package 'color.sty' is not loaded}%
    \renewcommand\color[2][]{}%
  }%
  \providecommand\transparent[1]{%
    \errmessage{(Inkscape) Transparency is used (non-zero) for the text in Inkscape, but the package 'transparent.sty' is not loaded}%
    \renewcommand\transparent[1]{}%
  }%
  \providecommand\rotatebox[2]{#2}%
  \newcommand*\fsize{\dimexpr\f@size pt\relax}%
  \newcommand*\lineheight[1]{\fontsize{\fsize}{#1\fsize}\selectfont}%
  \ifx\svgwidth\undefined%
    \setlength{\unitlength}{603.58744812bp}%
    \ifx\svgscale\undefined%
      \relax%
    \else%
      \setlength{\unitlength}{\unitlength * \real{\svgscale}}%
    \fi%
  \else%
    \setlength{\unitlength}{\svgwidth}%
  \fi%
  \global\let\svgwidth\undefined%
  \global\let\svgscale\undefined%
  \makeatother%
  \begin{picture}(1,0.57417196)%
    \lineheight{1}%
    \setlength\tabcolsep{0pt}%
    \put(0,0){\includegraphics[width=\unitlength,page=1]{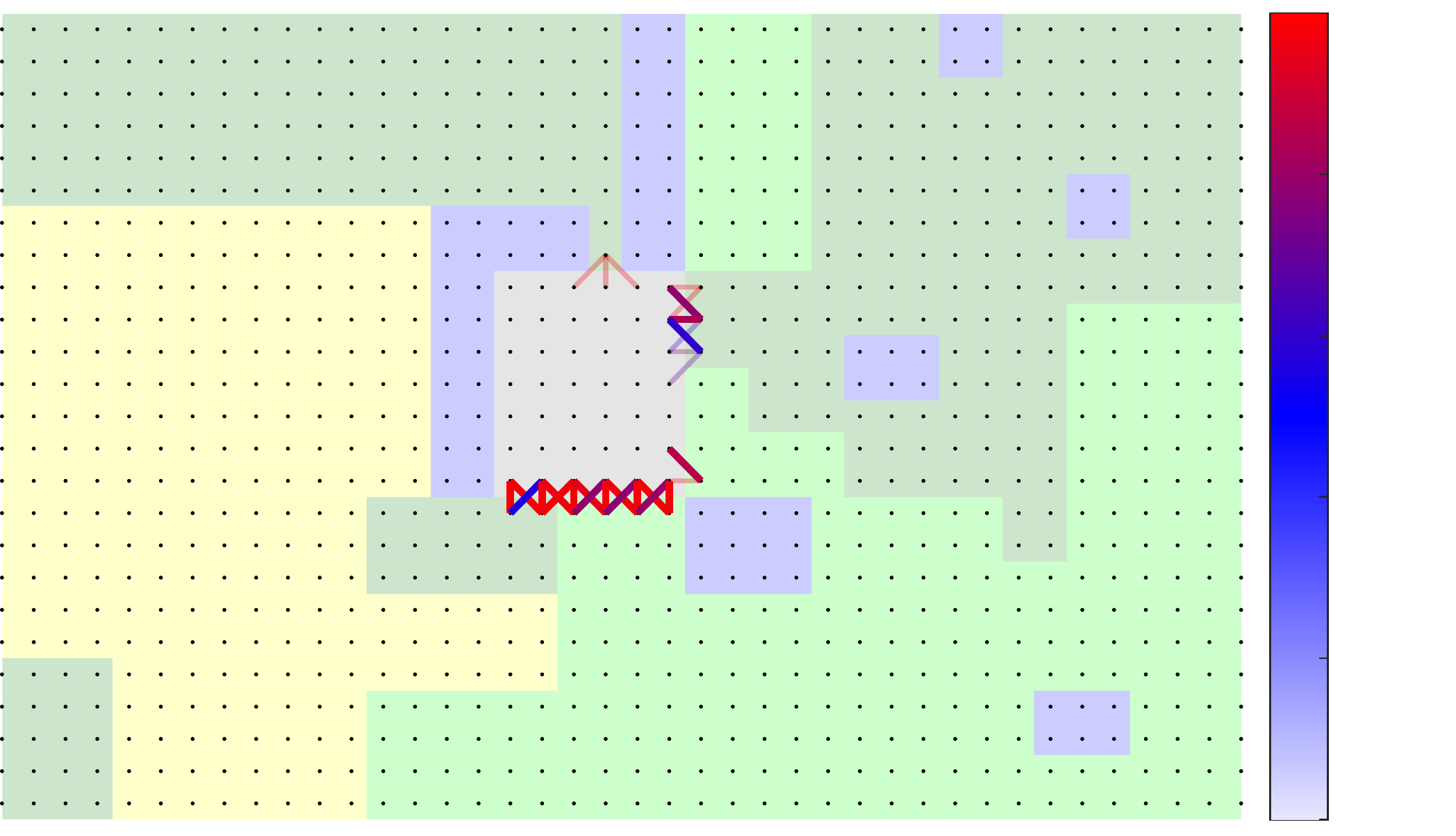}}%
    \put(0.92250924,0.00040626){\makebox(0,0)[lt]{\lineheight{1.25}\smash{\begin{tabular}[t]{l}\tiny 0\end{tabular}}}}%
    \put(0.92250924,0.11124356){\makebox(0,0)[lt]{\lineheight{1.25}\smash{\begin{tabular}[t]{l}\tiny .2\end{tabular}}}}%
    \put(0.92250924,0.22208085){\makebox(0,0)[lt]{\lineheight{1.25}\smash{\begin{tabular}[t]{l}\tiny .4\end{tabular}}}}%
    \put(0.92250924,0.33291815){\makebox(0,0)[lt]{\lineheight{1.25}\smash{\begin{tabular}[t]{l}\tiny .6\end{tabular}}}}%
    \put(0.92250924,0.44375544){\makebox(0,0)[lt]{\lineheight{1.25}\smash{\begin{tabular}[t]{l}\tiny .8\end{tabular}}}}%
    \put(0.92250924,0.55459274){\makebox(0,0)[lt]{\lineheight{1.25}\smash{\begin{tabular}[t]{l}\tiny 1\end{tabular}}}}%
    \put(0.99999,0.0809704){\rotatebox{90}{\makebox(0,0)[lt]{\lineheight{1.25}\smash{\begin{tabular}[t]{l}\scriptsize Resource allocation\end{tabular}}}}}%
    \put(0,0){\includegraphics[width=\unitlength,page=2]{Fire3.pdf}}%
  \end{picture}%
\endgroup%
 
\caption{k=3}
      \label{fig:032}
  \end{subfigure}
      \caption{Resource allocation for the scenario in Fig. \ref{fig:53} with $\Gamma^k=10$, $h=0.25$ and $\alpha=0.9$ for a forecast with no wind at k=1, a northeasterly wind of $V=4$ m/s at k=2, changing to a 8 m/s southerly wind midway step k=3.}     
\label{fig:FireRA}
\end{figure*}

We now assume the following weather forecast. At time step $k=1$; no wind, $k=2$ northeasterly wind of $V=4$ m/s, midway time step $k=3$ this changes to a southerly wind of $V=8$ m/s which is predicted to stay for at least another 2 time steps. Hence, in order to model this we can take the average graph of the two wind conditions for $k=3$.  Setting $\Gamma^k=10$, $h=0.25$, $\alpha=0.9$, we obtain the results illustrated in Fig. \ref{fig:FireRA}. The accumulated amount of resources are shown, where the new resources added per time step have a thicker line width and previous resources added a higher transparency. It can be seen that effectively a protective layer is created around the city following the forecast. First, with no wind and northeasterly wind coming up, resources are allocated to the north and east, afterwards the city is prepared for the incoming southerly wind with more resources allocated to the south to protect the city from higher spread from those areas.  

Scalability of the proposed convex optimization program was investigated in \cite{TCNS2021,ACC2022} and solver time was found to be linear in both number of nodes and time-steps. For the dynamic variation, similar run times were found. 


%

\bibliography{ACFR2021}             

\end{document}